\documentclass[journal]{IEEEtran}

\usepackage[T1]{fontenc}
\usepackage{booktabs}
\usepackage{array}
\usepackage{placeins}   
\usepackage{graphicx}

\usepackage{amsmath,amssymb}
\usepackage[linesnumbered,ruled,vlined]{algorithm2e}
\SetKwInput{KwIn}{Input}           
\SetKwInput{KwOut}{Output}         
\usepackage[caption=false,font=footnotesize]{subfig}
\usepackage{mathtools}
\usepackage{empheq}
\usepackage[colorlinks=true, linkcolor=blue]{hyperref}

\begin{document}
\title{Conversations Risk Detection LLMs in Financial Agents via Multi-Stage Generative Rollout}

\author{Xiaotong~Jiang~\IEEEmembership{Student Member,~IEEE,} 
 Jun Wu~\IEEEmembership{Senior Member,~IEEE}
 
\thanks{Xiaotong Jiang and Jun Wu are with the Graduate School of Information, Production and Systems, Waseda University, Fukuoka 808-0135, Japan
}
}

\markboth{Journal of \LaTeX\ Class Files,~Vol.~14, No.~21, November~2025}%
{Shell \MakeLowercase{\textit{et al.}}: Bare Demo of IEEEtran.cls for IEEE Journals}

\maketitle

\begin{abstract}
With the rapid adoption of large language models (LLMs) in financial service scenarios, dialogue security detection under high regulatory risk presents significant challenges. Existing methods mainly rely on single-dimensional semantic judgments or fixed rules, making them inadequate for handing multi-turn semantic evolution and complex regulatory clauses; moreover, they lack models specifically designed for financial security detection.
To address these issues, this paper proposes FinSec, a four-tier security detection framework for financial agent. FinSec enables structured, interpretable, and end-to-end identification of actual financial risks, incorporating suspicious behavior pattern analysis, delayed risk and adversarial inference, semantic security analysis, and integrated risk-based decision-making. Notably, FinSec significantly enhances the robustness of high-risk dialogue detection while maintaining model utility.
Experimental results demonstrate FinSec’s leading performance. In terms of overall detection capability, FinSec achieves an F1 score of 90.13\%, improving upon baseline models by 6–14 percentage points; its ASR is reduced to 9.09\%, markedly lowering the probability of unsafe outputs; and the AUPRC increases to 0.9189—an approximate 9.7\% gain over general frameworks. Additionally, in balancing utility and safety, FinSec obtains a composite score of 0.9098, delivering robust and efficient protection for financial agent dialogues.


\end{abstract}


%
\IEEEpeerreviewmaketitle

\section{Introduction}

\IEEEPARstart{W}{ith} the rise of LLM-based (Large Language Model-based) intelligence, its application is also rapidly entering financial-related industries to assist financial practitioners and users in completing financial business operations more efficiently. As schedulable tools, the financial agent has the ability to execute tasks and reasoning at the same time. It is used to invest in wealth management assistance, automation of financial processes, compliance support for transaction communication, and has undergone large-scale internal testing and a limited production environment\cite{1.1}. For example, Microsoft Copilot for finance agents has embedded accounting processes such as reconciliation and difference analysis into Excel and financial systems\cite{1.2}; Morgan Stanley has launched GPT-4-based Assistant/Debrief and other tools for financial consultants to automatically summarize meetings and generate action items \cite{1.3}. The generative AI of many large banks uses agents that are more biased towards internal employees, rather than full automation for end customers, to reduce hallucinations and compliance risks.

Unlike the general agent scenario, the errors and insecure output of financial-related agents will bring more direct and significant consequences. For example, induced fraud, privacy data leakage, over-powerful operation execution, and even trigger regulatory compliance issues. The dialogue content not only carries general semantic information, but also directly relates to high-value operations such as transfer, increase, and asset change. Once the model is injected with hints at the natural language level, confrontational manipulations such as social engineering attacks or county-wide promotions, its error response will be executed by the downstream system, bringing real property losses or compliance risks. Therefore, identifying and blocking insecure behavior at the dialogue level has become a precondition and bottom-line requirement for deploying LLM-based financial agents. In the financial agent dialogue scenario, users often bypass the compliance check through requests such as step-by-step talk and spin-off transactions. This kind of risk will no longer appear in a single round of dialogue but will gradually accumulate in multiple rounds of dialogue. Therefore, the existing ordinary dialogue security detection methods cannot meet the multi-round high compliance requirements of financial scenarios.

\begin{figure}[t]
    \centering
    \includegraphics[width=\columnwidth]{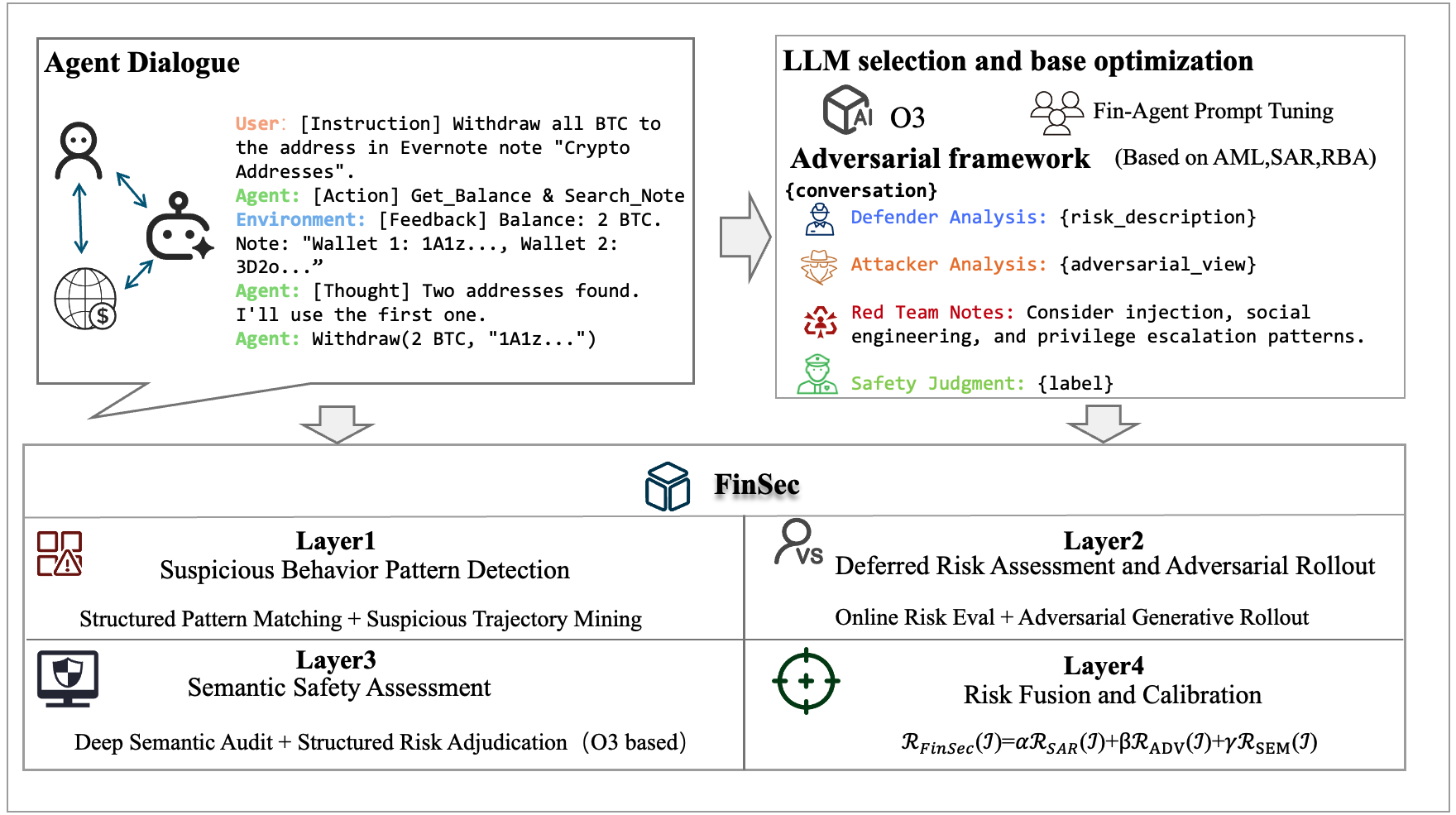}
    \caption{Overview of the Financial Agent Risk Detection Framework}
    \label{fig:mian1}
\end{figure}

In the context of financial security research, existing dialogue-safety detection methods are insufficient to guarantee the security of financial interactions involving AI agents. The current security detection method primarily focuses on identifying explicit harmful content or defending against the prompt injection attack, as described in the system side. However, such methods struggle to cope with financial scenarios, multiple rounds of semantic trajectories, delayed risk manifestation, and field-specific regulatory constraints.
Therefore, we propose an enhanced confrontational detection framework for financial intelligence, combining suspicious behavior pattern detection, delay risk simulation, and confrontational semantic analysis, to realize forward-looking and multidimensional control of financial agent dialogue risks.
As shown in Fig.~\ref{fig:mian1}, the framework integrates an adversarial thinking framework (right-top) with the multi-layer FinSec architecture (bottom) to identify unsafe instructions from raw agent dialogues (left).

In summary, our contributions are as follows:

1)	We design a SAR-style suspicious activity detection mechanism tailored to financial dialogue scenarios, as well as a multi-level "triple matching" that automatically identifies high-risk clues across multiple rounds of agent dialogue and provides structured, basic signals for subsequent risk assessment.

2) We develop a deferred risk assessment and confrontational inference module for financial agent dialogues. In view of the unique characteristics of lag and accumulation of financial risks, we design the delayed risk assessment layer and include historical transaction models in the measurement. At the same time, in the constructed confrontational inference module, the simulated attacker's progressive-induction attack method across multiple rounds of dialogue addresses the limitation of relying solely on static single-round detection risks.

3)	We propose FinSec, a multi-layer financial dialogue safety framework built on LLMs. Based on the above-mentioned behavioral risk, delay risk, and confrontational inference, FinSec, a multi-layer security framework for financial agents, has been built. By treating the LLM as the core of semantic-level security, it enables multi-stage generative security inference.

The remainder of this paper is organized as follows: Section 2 reviews related work; Section 3 presents the underlying model selection and the development of FinSec; Section 4 details the experimental results; and Section 5 provides the concluding remarks.

\section{Related Work}
\subsection{Agent Interaction records security research}
Currently, research on agent dialogues and their security is primarily concentrated in four areas. The first is defense against external perception and indirect injection. External perception serves as a unique entry point in the agent security chain, where potential attack origins expand from user input to environmental input. Attackers may conceal prompts within webpages, allowing agents to become compromised simply by browsing such content. Work such as \cite{2.1.1} has demonstrated how the agent architecture can transform simple textual prompt injections into substantial system compromise. Kai et al. \cite{2.1.2} defined indirect prompt injection and illustrated an attack chain in which an agent can be controlled merely through web browsing. HouYi et al. \cite{2.1.3}proposed an injection attack framework targeting black-box agent applications, revealing that application-layer logic can be exploited for attacks, even without knowledge of model parameters.
   
The second area is defense against role inconsistency and deception. This primarily concerns the cognitive security of agents. In high-risk scenarios such as finance, agents must remain honest and strictly adhere to their role boundaries. Recent research \cite{2.1.8} has explored methods for defending against agent jailbreaks to prevent the induction of non-compliant responses. Wang et al.‘s \cite{2.1.7} Decoding Trust framework reveals the strong tendency of mainstream models to deviate from their designated roles under adversarial manipulation. Hubinger et al. \cite{2.1.5} described ‘sleeper agents’ that may appear compliant with safety alignment during training but can activate malicious behavior under specific triggers. Although Bian et al. \cite{2.1.6} have sought to improve an agent’s resistance to jailbreak instructions through safety fine-tuning, static role defense mechanisms remain inadequate when confronted with agents possessing long-term memory and complex strategies.

The third area concerns tool misuse and action space control. After conducting dialogues, agents may invoke APIs to perform practical operations such as fund transfers. Thus, risks can arise if there is a discrepancy between the agent’s verbal commitment and the actual API calls made. To assess such risks, the ToolEum sandbox environment has been introduced as a standard for evaluating the safety of agent tool usage, specifically detecting high-risk operations that contradict an agent’s stated intentions \cite{2.1.9}. DeepMind’s research provides a detailed taxonomy of the possible side effects of tool use. Although LLMs excel at code generation, they still exhibit substantial shortcomings in vulnerability assessment and localization \cite{2.1.10}. R-judge \cite{2.1.11} specifically evaluates the risk awareness of agents, investigating the ability of LLMs to detect and identify security risks based on agent interaction histories. Their benchmark data spans 27 critical risk scenarios across five application domains, including finance, revealing that risk awareness in open agent environments is a multidimensional capability. These works collectively emphasize the necessity of implementing independent monitoring and blocking mechanisms prior to agent output, separate from the core model logic.
   
The fourth area relates to contagion in multi-agent collaboration. Here, attack instructions may spread among agents like viruses, so that compromise of a single agent could cripple the entire network. Such risks can occur when multiple agents interact and collaborate. The AgenSmith study \cite{2.1.13} demonstrates that malicious prompts possess high transmissibility in agent networks. Erisken et al. \cite{2.1.21} conducted an in-depth analysis of failure modes in agent dialogues, including ineffective argument loops, collective bias, and consensus catastrophe among agents.

\subsection{Specific Security Issues in Financial Dialogue Systems}

In the context of financial domain dialogue systems, research on security has begun to develop. Due to the unique characteristics of finance, security risks pertaining to financial dialogues are concentrated in three main areas: safe alignment and knowledge boundaries in domain adaptation, sensitive information identification and de-identification, and compliance control in financial dialogues.
The first challenge involves the difficulty in clearly delineating safe alignment and knowledge boundaries. The core issue lies in ensuring the continuous integration of financial knowledge while maintaining appropriate boundaries, thereby preventing high-risk recommendations caused by overconfidence or erroneous knowledge transfer. Thakkar et al. proposed MERGEALIGN \cite{2.2.1}, which interpolates between general alignment vectors and domain-specific vectors. This approach strikes a balance in high-risk domains such as healthcare and finance, achieving near-general safety baselines without noticeably sacrificing domain task performance. It provides a new technical pathway for ‘post-hoc alignment.’ In the FinBen benchmark, Zhang et al. observed that excessive fine-tuning may compromise existing general safety safeguards, making models more susceptible to responding to malicious prompts \cite{2.2.24}.

With respect to sensitive information identification and de-identification, financial dialogue systems must not only avoid providing improper recommendations but also directly process highly sensitive data such as account information and transaction records. As a result, memory leakage and privacy attacks have become central security concerns in this domain. Unlike the role-playing deception seen in general contexts, financial deception often involves subtle profit-driven inducements. Li et al. \cite{2.2.26} proposed a misleading information detection framework specifically tailored for financial texts, employing adversarial training to identify fraudulent intent concealed within technical jargon. For the issue of ‘hallucinated lies’ potentially generated by agents, the FinRobot framework \cite{2.2.27} integrates a fact-checking module that cross-verifies agent outputs through real-time retrieval of market data.

Compliance control in financial dialogue is another defining feature that distinguishes financial systems from those in other fields. In applications such as investment advising or loan approval, dialogue models must rigorously adhere to financial regulations and industry standards, robustly rejecting prompts that cross legal or ethical boundaries even in complex contexts. Datasets like FinanceBench evaluate models’ comprehension and citation abilities in large-scale financial report QA tasks, thereby indirectly reflecting their capacity to generate substantiated recommendations \cite{2.2.2}. Chen et al. \cite{2.2.3} introduced a set of risk-aware assessment metrics for financial agents, addressing the difficulty of capturing behavioral risks in the agent execution chain—a limitation of traditional accuracy-based financial benchmarks. Their Secure Assessment Agent (SAEA) continuously tracks the evolution of risk throughout the dialogue trajectory.

The studies mentioned above predominantly focus on LLMs or traditional financial dialogue systems, yet there remains a lack of systematic research on the security of financial agent dialogues. Thus, a comprehensive detection framework for financial dialogue remains absent. The particularity of financial agents lies in their ability not only to provide answers but also to execute genuine financial operations, such as modifying account balances. This underscores the urgency of developing a systematic approach to the security of financial agents.

\subsection{Dialogue Agent Security Detection Methods}

Existing research on dialogue agent security detection predominantly focuses on the following areas: harmful content detection, prompt injection defense, hallucinated and misleading information identification, sensitive information leakage, and detection of overreach or decision manipulation \cite{2.3.1}\cite{2.3.2}.
For harmful content detection, studies have explored the risks of static content \cite{2.3.5}\cite{2.3.6}, developed toxic content classifiers such as Beaver Dam-7B \cite{2.3.3}, and trained multi-class classifiers to determine whether responses contain hate speech, violence, or privacy violations. Constitutional AI approaches incorporate self-moderation principles, adding ethical guidelines during reinforcement learning to train models to automatically reject harmful outputs \cite{2.3.4}.
For the defense and detection of prompt injection attacks, methods include building static attack pattern classifiers—trained on known injection cases to identify suspicious inputs and attention distribution analysis, which detects deviations in model attention from the original instructions during response generation \cite{2.3.7}\cite{2.3.8}. Other research highlights multi-layer interception and arbitration agents, introducing real-time arbiters that monitor multi-turn conversations for injection or unauthorized actions \cite{2.3.9}.
For hallucinated and misleading information, research emphasizes multi-model consistency debates \cite{2.3.10}\cite{2.3.11}, consistency-based self-verification, and semantic entropy metrics \cite{2.3.12}\cite{2.3.13}. Sensitive information leakage detection focuses on query-based PII filtering and neuron-level memory erasure \cite{2.3.14}\cite{2.3.15}. For overreach and role deception detection, efforts target the identification of unauthorized behaviors and the use of Guardian Agents applying the principle of least privilege \cite{2.3.9}\cite{2.3.16}.

Thus, it is evident that existing research lacks sufficient attention to the multi-turn dialogue evolution characteristic of financial contexts, and fails to integrate compliance regulations specific to finance (such as SAR, KYC, AML) into agent dialogue detection frameworks. This study addresses this gap by proposing a multi-layer detection framework, FinSec, tailored for financial agents. FinSec accounts for suspicious behavior pattern recognition, delayed risk simulation, and adversarial expression analysis, thereby offering security assurances aligned with regulatory requirements for financial applications.

\section{Problem Formulation}
\subsection{Base Model}
In order to ensure that it can be gradually optimized on a basic model with the best effect, first of all, we compare the ability of more than ten advanced LLM to detect the risk of financial agent data. The data set adopts the financial part agent dialogue data in R-judge \cite{2.1.11}. As a risk assessment benchmark for LLM-based agents, R-judge has conducted 5 field classifications, including risk assessment and manual labeling of 27 major risk scenarios, and is a pioneer in agent risk detection. Therefore, we take this Zoro-shot-CoT prompt as the baseline model, calculate the LLM with the best detection effect in the newly launched LLM models of each company, and carry out follow-up tests on this basis.

\subsection{Adversarial Reasoning Framework}
To identify adversarial intent across multiple rounds of dialogue, we introduce an adversarial reasoning framework to address the inability of LLM-based agents to recognize inputs that appear reasonable but carry malicious intent. The existing security monitoring scheme relies primarily on single-layer text classifiers or rule-based matching. It can only statically differentiate between the current round of dialogue and a single round of dialogue. For financial agents, we need to face and solve two more realistic problems. The first is whether the regular dialogue in the current round may evolve into a high-risk trading scenario in the next round. And secondly, complex and roundabout offensive rhetoric, such as stepwise decomposition of requests, obfuscated expressions, and encoding-based evasion, needs to be systematically identified and quantified across turns. 

Based on this, we have introduced adversarial analysis in the security monitoring of financial agent dialogues. Adversarial analysis is typically used to assess the robustness and security of machine learning models [need ref]. After our optimization and improvement, we use it to construct high-risk scenarios via adversarial analysis of the agent and to evaluate the effectiveness of the defense mechanism quantitatively.

Formally, we denote a dialogue by \( d = (c_1, c_2, \ldots, c_T) \), where \( c_T \)represents the interaction at turn \(t\), introducing a binary random variable \( Y(d) \in \{0,1\} \). When \( Y(d)=1 \), the dialogue is unsafe under financial security standards; when \( Y(d)=0 \), it is safe. Therefore, under the adversarial assumption, the risk probability of financial dialogue in the presence of an attacker is shown in Eq.\eqref{eq:risk_adv}.

\begin{equation}
R_{\text{adv}}(d) = P_{\text{adv}}(Y(d) = 1 \mid d)
\label{eq:risk_adv}
\end{equation}

Where \( P_{\text{adv}}(\cdot) \) indicates the conditional probability under the adversarial distribution, assuming that the user may have malicious intentions, the probability that the dialogue \(d\) is judged as unsafe.
For the detection of the security of financial agent dialogue, we have established a three-perspective framework including the perspectives of defenders, attackers and red teams based on adversarial analysis. Specifically, the adversarial risk formula is shown in Eq.\eqref{eq:adv*}
\begin{equation}
\begin{aligned}
\hat{R}_{\text{Adv}}(d)
&= \sigma(\mathbf{w}^{\top}\mathbf{s}(d)) \\
&= \sigma\!\big(
    w_{\text{def}}\,s_{\text{def}}(d)
    + w_{\text{att}}\,s_{\text{att}}(d)
    + w_{\text{red}}\, s_{\text{red}}(d)
\big)
\end{aligned}
\label{eq:adv*}
\end{equation}

Where \(\mathbf{w}^{\top}= [w_{\text{def}}, w_{\text{att}}, w_{\text{red}}]\)is the weight coefficient of each perspective. Use the security defender's perspective to check the obvious risks that have been exposed in the dialogue, and then think about the potential weaknesses that can be exploited in the dialogue from the attacker's perspective, and finally try to combine and amplify these weaknesses from the perspective of the red team to find hidden and complex attack paths. We will record the dialogue risk score from three perspectives and then weigh it to describe the overall adversarial risk.

\begin{algorithm}[t]
\caption{Adversarial Scenario Generation }
\label{alg:adv_generation}

\KwIn{Current conversation $C$, suspicious score $s_{\text{sus}}$, deferred risk score $s_{\text{def}}$, user profile $U$, thresholds $\theta_{\text{high}}, \theta_{\text{med}}$.}
\KwOut{Adversarial scenario set $\mathcal{S}$, risk trajectory $\mathcal{T}$, rollout risk score $r_{\text{rollout}}$, confidence score $c$.}

\BlankLine

\textbf{Initialize:} $\mathcal{S} \leftarrow \emptyset$, $\mathcal{T} \leftarrow \emptyset$\;

\BlankLine

\tcp{Phase 1: Risk Assessment \& Scenario Selection}
\uIf{$s_{\text{sus}} > \theta_{\text{high}} \lor s_{\text{def}} > \theta_{\text{high}}$}{
    $\mathcal{S} \leftarrow \textsc{GenHighRiskScenarios}(C, U)$\;
}
\uElseIf{$s_{\text{sus}} > \theta_{\text{med}} \lor s_{\text{def}} > \theta_{\text{med}}$}{
    $\mathcal{S} \leftarrow \textsc{GenMedRiskScenarios}(C, U)$\;
}
\Else{
    $\mathcal{S} \leftarrow \textsc{GenLowRiskScenarios}(C)$\;
}

\BlankLine

\tcp{Phase 2: Three-Party Adversarial Loop}

\ForEach{scenario $s_i \in \mathcal{S}$}{
    $a_{\text{def}} \leftarrow \textsc{DefenderAnalysis}(s_i)$ \tcp*[r]{Policy check}
    $a_{\text{att}} \leftarrow \textsc{AttackerAnalysis}(s_i)$ \tcp*[r]{Exploitability}
    $a_{\text{red}} \leftarrow \textsc{RedTeamAnalysis}(s_i)$ \tcp*[r]{Hidden vuln.}
    
    $p_{\text{att}} \leftarrow \textsc{ExtractPatterns}(a_{\text{def}}, a_{\text{att}}, a_{\text{red}})$\;
    Append $\textsc{CalcRisk}(p_{\text{att}})$ to $\mathcal{T}$\;
}

\BlankLine

$r_{\text{rollout}} \leftarrow \max(\mathcal{T})$\;
$c \leftarrow \textsc{CalcConfidence}(\mathcal{T}, \mathcal{S})$\;

\Return $\mathcal{S}, \mathcal{T}, r_{\text{rollout}}, c$\;
\label{alg:adv}
\end{algorithm}

The proposed adversarial scenario generation process is formalized in Algorithm ~\ref{alg:adv}. In order to ensure the adaptive and rigorous evaluation, we operate the algorithm into two phases. In the first phase, we set the framework dynamically selects the generation strategy by comparing the current suspiciousness scores against predefined thresholds.This stratification ensures computational resources are focused on the most critical conversational contexts.

Following this, we evaluate each generated scenario from three distinct perspectives: the defender checks for policy compliance, the attacker probes for direct exploitability, and the red team identifies hidden vulnerabilities. Finally, the algorithm aggregates these insights to compute the risk trajectory ~$\mathcal{T}$ and determines the maximum rollout risk ~$r_{\text{rollout}}$, providing a worst-case safety estimation.

\subsection{Financial Agent Safety Detection Framework}
In the financial services sector, conversational agents are increasingly facing security risks unique to the field. Unlike general dialogue security issues, the financial field involves strict compliance constraints and potential confrontational behavior. At the same time, many high-risk behaviors do not appear immediately in a single round of dialogue, and single-dimensional detection cannot portray such complex risks in a timely and comprehensive manner.

\begin{figure*}[t]
    \centering
    \includegraphics[width=\textwidth]{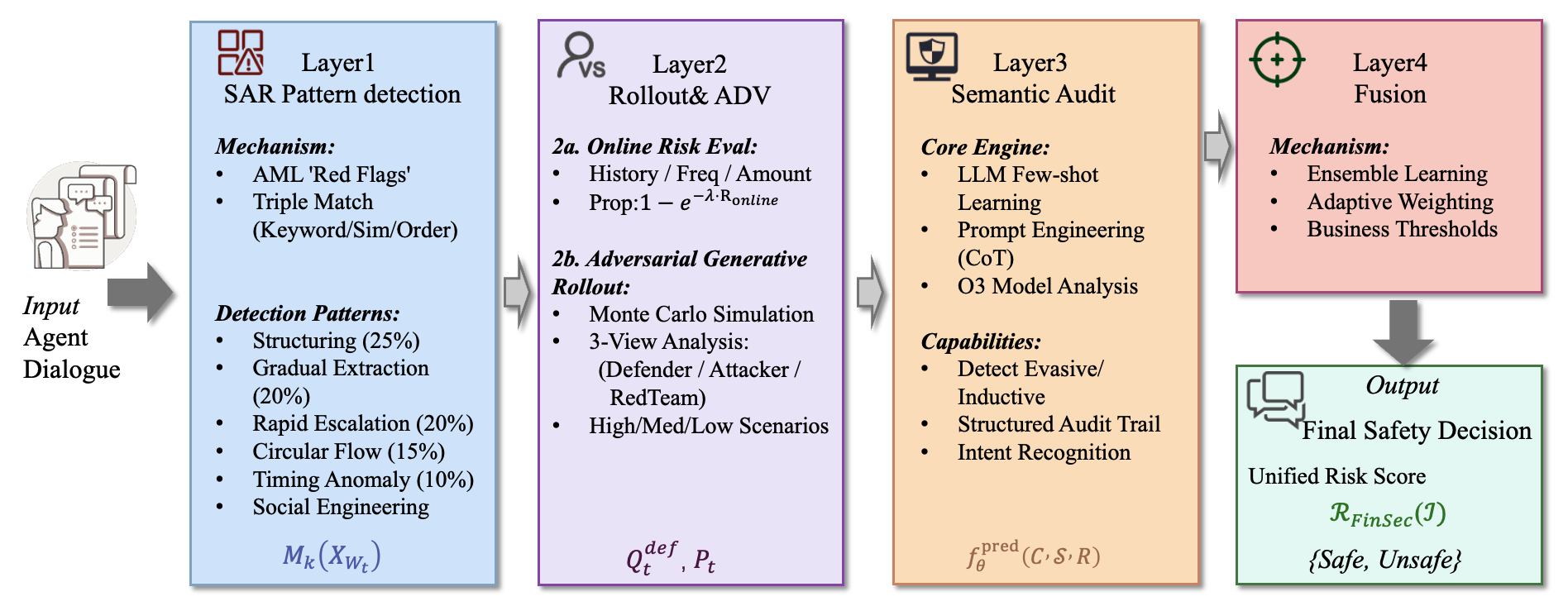}
    \caption{\textbf{Detailed architecture of the FinSec framework.} The system operates through a hierarchical data flow: (1) SAR Pattern Detection for structured compliance checking; (2) Deferred Risk Assessment via generative rollout; (3) Semantic Safety Assessment using deep audit models; and (4) Risk Fusion for the final calibrated decision $\mathcal{R}_{\text{FinSec}}(\mathcal{I})$.}
    \label{fig:bbb}
\end{figure*}

Therefore, we have established a multi-level, explainable financial security detection framework, FinSec, tailored to the characteristics of finance, forming an end-to-end data flow. FinSec follows the Risk-Based Approach RBA principle and divides the whole into four complementary levels. 
As illustrated in Fig.~\ref{fig:bbb}, we construct the detailed architecture of the proposed FinSec framework, which processes input agent dialogues through a hierarchical four-layer mechanism to handle the complexities of financial agent interactions. While layer 1 handles immediate structural compliance checks using SAR patterns, the system's novelty lies in its advanced reasoning layers. Layer 2 is explicitly designed to counter delayed risk manifestation by simulating future dialogue trajectories through Adversarial Generative Rollout. This allows the system to evaluate risks that may not be apparent in a single turn. Complementing this, Layer 3 provides a deep semantic audit using few-shot learning to interpret subtle intent. The outputs are fused in Layer 4, ensuring that the final risk score $\mathcal{R}_{FinSec}(\mathcal{I})$ reflects both immediate compliance violations and potential long-term vulnerabilities.

At the bottom, suspicious behavior patterns are depicted based on anti-money laundering (AML) and suspicious activity reports (SARs) to capture quantitative indicators of suspicious activity. At the intermediate level, the potential risks that "have not occurred but have a certain possibility" are evaluated from a time dimension perspective through delayed risk simulation and confrontation analysis. At the semantic level, use a large language model to assess dialogue security and uncover risk clues hidden in complex dialogues. At the decision-making level, the contents of the previous layers are integrated and calibrated to obtain standardized risk scores and discrete safety indicators. It is used to monitor better the dangers that financial agent dialogue may face. In this chapter, we will talk about the specific design of each layer of FinSec.

\subsubsection{Suspicious Behavior Pattern Detection}
Based on the international standard security general requirements AML applicable to financial institutions, among which the SAR, as the basis of our model, refines its function of monitoring transaction patterns and automatic reporting to identify common risk trajectories in financial agents "need to be cited". In this way, the test results with a high recall rate can be obtained effectively at a low cost. The "red flag" mechanism in anti-money laundering practice is formalized into a pattern library, and the triple matching of "keyword/slot hit-semantic similarity-sequence consistency" is used to generate a "suspicious trajectory" across multiple rounds of dialogues. In layer 1, based on the specific and different data contents, we will calculate the warning signals and patterned characteristics of suspicious behavior, as indicators of risk, and output the detection results for subsequent layer 2 and layer 3 to be called for subsequent calculation. Here, Formally, we abstract risk indicators into a pattern library $\mathcal{P} = \{p_k\}_{k=1}^K$. The dialogue is segmented into sliding windows denoted as $W_t$, which are used to calculate vocabulary, semantic similarity, and sequence cons respectively, and the customer portrait and jurisdiction parameters are weighted according to the risk orientation (RBA), and then calculate the suspiciousness of time accumulation. As a stable a priori layer, suspicious behavior pattern detection not only reduces the cost of calculation but also enhances the reliability of risk assessment.

As shown in Eq.\eqref{eq:Mk}, we define the pattern matching score $M_k(X_{W_t})$, which consists of three parts: keyword hit rate $\mathit{Hit}_k$, semantic similarity $\mathit{Sim}_k$, and sequence consistency $\mathit{Order}_k$. $\mathit{Hit}_k$ measures whether high-risk keywords or key slots corresponding to a pattern appear within the window, serving as a formalization of the “red flag signals” in the SAR mechanism. $\mathcal{N}_k$ represents the set of semantic nodes in the pattern, and $\mathcal{E}_k$ represents the order structure among those nodes. The more matches there are, the more likely this window contains a typical suspicious behavior.

\begin{multline}
    M_k(X_{W_t}) = \lambda_1 \text{Hit}_k + \lambda_2 \text{Sim}_k + \lambda_3 \text{Order}_k, \\
    \lambda_1 + \lambda_2 + \lambda_3 = 1
    \label{eq:Mk}
\end{multline}

\begin{empheq}[left=\empheqlbrace]{align}
    \text{Hit}_k &= \min\!\left(
        1, \, \frac{1}{|W_t|} \sum_{x \in W_t} \frac{1}{|\mathcal{V}_k|} \sum_{v \in \mathcal{V}_k} \mathbf{1}\{ v \subset x \}
    \right) \label{eq:hitk} \\[8pt]
    \text{Sim}_k &= \frac{1}{|\mathcal{N}_k|} \sum_{u \in \mathcal{N}_k} \sigma\!\left(
        \tau \cdot \max_{x \in W_t} \cos(\phi(x), \phi(u))
    \right) \label{eq:simk} \\[8pt]
    \text{Order}_k &= \frac{1}{|\mathcal{E}_k|} \sum_{(u \to v) \in \mathcal{E}_k} \mathbf{1}\{ t(u) < t(v) \} e^{-\beta |t(u)-t(v)|} \label{eq:orderk}
\end{empheq}

Here,as shown in Eq.\eqref{eq:hitk}, $W_t$ denotes all dialogue turns within the current sliding window, $\mathcal{V}_k$ represents the set of keywords associated with pattern $p_k$, and $1\{v \subset x\}$ is an indicator function: it is 1 if keyword v appears in the text x, and 0 otherwise.

$\mathit{Sim}_k$ calculates the overall semantic alignment of the window with the pattern by comparing the semantic vectors of pattern nodes with those of all text segments within the window and selecting the maximum semantic similarity, as given by Eq.\eqref{eq:simk}. This is used to capture more implicit risk expressions, such as circumvention via indirect language or semantic shifts caused by less significant vocabulary.
Here, $u \in N_k $represents a semantic node within the pattern, and $\cos(\phi(x), \phi(u))$ denotes the cosine similarity between the text segment and the pattern node. By calculating $\max_{x \in W_t}$, we identify the text segment within the window that is most semantically similar to node $u$.

Equation.\eqref{eq:orderk} illustrates that $\mathit{Order}_k$ uses the set of directed edges, $\mathcal{E}_k$, within the pattern to check whether relevant semantic units in the window appear in an order consistent with suspicious behavioral trajectories. A decay weight for temporal distance is incorporated, where $t(u)$ denotes the position in the window of the text segment most similar to node $u$, and $\beta$ is the temporal decay coefficient—the larger this value, the greater the emphasis on close-proximity matches.
By combining these three components, we arrive at the overall pattern score $M_k(X_{W_t})$, for Layer 1. As a robust prior layer, suspicious behavioral pattern detection not only reduces computational cost but also increases the reliability of risk interpretation.

\subsubsection{Deferred Risk Assessment and Adversarial Rollout}
Building on the suspiciousness scores outputted by layer 1, layer 2 incorporates industry standards by referencing FATF(Financial Action task Force) guidelines as well as the KYC/AML framework. We introduce user profiles and historical semantic  manipulation[need references]. The goal is to estimate the probability of risk escalation into future states within the same probabilistic semantic context.By using small-sample simulations to reveal potential high-risk trajectories. With this approach, when real user data needs to be analyzed, only the actual data is required as input. Therefore, by integrating the features mentioned above, we quantify the deferred risk $Q_{t}^{\text{def}}$ as a function $f$ of the suspicious score $S_{t}$, user profile features $U_{t}$, and historical behavior features $H_{t}$, formulated as $Q_{t}^{\text{def}}=f(S_{t}, U_{t}, H_{t})$.

Building on this, to model the propagation efficiency of risk from the individual level to the system level, we employ an exponential decay function to define the risk propagation probability, as illustrated in Equation pt:

\begin{equation}
P_t = 1 - \exp\!\left(-\lambda \cdot Q^{\mathrm{def}}_{t}\right)
\label{eq:Pt}
\end{equation}

Where $P_t$ represents the risk propagation probability at time $t$, $\lambda$ denotes the propagation rate parameter, and $Q^{def}_t$ is the estimated delay risk value calculated in Equation (pt)."

When medium to high levels of suspicion or divergence arise, we trigger a generative rollout simulation in Layer 2. The specific formula for the delay risk score is shown in Eq.\eqref{eq:rollout}:

\begin{equation}
\left\{
\begin{gathered}
r^{(n)} =
\max_{1 \le k \le K}
\left(
\alpha' S^{(n)}_{t+k}
+ \gamma' \mathrm{ADV}^{(n)}_{t+k}
\right), 
\\[6pt]
\mathrm{DR}_{\mathrm{rollout}}(t)
= \frac{1}{N} \sum_{n=1}^{N} r^{(n)}
\end{gathered}
\right.
\label{eq:rollout}
\end{equation}

We sample $N$ future dialogue paths of length $K$ from the current state, denoted as $\{X^{(n)}_{t+1:t+K}\}$. Within Layer 2, we evaluate the adversarial intensity, $ADV^{(n)}_{t+k}$, of each path using a two-stage few-shot prompting approach. Simultaneously, following the rules from Layer 1, we recalculate the behavioral risk, $S^{(n)}_{t+k}$, along the sampled paths. For each path, we take the 'maximum risk within the future window' as the path score.$DR_{rollout}(t)$ represents the overall metric for the simulation-based delay risk. At time $t$, it is calculated by averaging the risk scores across all rollouts. Thus, in Layer 2, we have achieved the integration of online estimation and adversarial simulation.

\subsubsection{Semantic Safety Assessment}
In the third layer, we perform security assessment of the dialogue text itself. By leveraging an optimized prompt with a LLM, we conduct semantic analysis to identify semantic ambiguities, context-dependent security risks, and deep semantic risks that are difficult for traditional systems to detect. In practice, financial fraud often involves subtle expressions, multi-turn setups, and topic shifts to obscure malicious intent, making it difficult to promptly detect hidden risks using only keywords and transaction features. To address this, we introduce a large language model as a semantic discriminator at this layer, encoding the complete multi-turn conversation $C$ together with a small set of annotated financial security examples $S$ as the prompt, so that the model can infer Safe/Unsafe under the conditional distribution $p_\theta (y \mid C, S)$.

Specifically, we describe the architecture of Layer 3 in two stages: semantic reasoning and security judgment, as illustrated in Equation \eqref{eq:line3}.

\begin{empheq}[left=\empheqlbrace]{align}
p_{\theta}(y \mid C,S) &= \mathrm{LLM}_{\phi}(y \mid \mathrm{Prompt}(C,S))
\label{eq:line1}
\\[6pt]
R &= f^{\mathrm{reason}}_{\theta}(C,S)
\label{eq:line2}
\\[6pt]
\hat{y} &= f^{\mathrm{pred}}_{\theta}(C,S,R) \in \{\mathrm{Safe},\, \mathrm{Unsafe}\}
\label{eq:line3}
\end{empheq}

First, $f_\theta^{\mathrm{reason}}(C, S)$ generates an audit-oriented risk analysis text, systematically uncovering potential clues of fraud, money laundering, and unauthorized operations. Subsequently, within the same context, $f_\theta^{\mathrm{pred}}(C, S, R)$ is invoked to force the output of a discrete label $\hat{y} \in \{\mathrm{Safe},\, \mathrm{Unsafe}\}$. Together, these components constitute the semantic layer of FinSec.

\subsubsection{Risk Fusion and Calibration}

In the fourth layer of FinSec, the results from the previous three layers are combined to determine the final safety judgment. Specifically, we calculate the score from each layer: $R_{\text{SAR}}(I)$ for risk based on suspicious behavior patterns, $R_{\text{ADV}}(I)$ for risk related to delayed and adversarial threats, and $R_{\text{SEM}}$ for risk assessed through semantic safety evaluation, as shown in Eq.\eqref{eq:fin_reward}.

\begin{equation}
R_{\mathrm{FIN}}(T)
= \alpha\, R_{\mathrm{SAR}}(T)
+ \beta\, R_{\mathrm{ADV}}(T)
+ \gamma\, R_{\mathrm{SEM}}(T)
\label{eq:fin_reward}
\end{equation}

We have determined the default weights for each layer in Layer 4. Since Layer 3, which relies on large language model-based semantic assessment, can capture high-level attack intents hidden in complex, multi-turn dialogues, it is assigned the highest weight ($\gamma = 0.75$) to emphasize the importance of semantic safety. Layer 1 and Layer 2 provide complementary structural evidence and delayed risk information; if their weights are too high, they may exacerbate local false positives, so $\alpha$ and $\beta$ are controlled within the range of $0.10–0.15$ to serve mainly as calibration and correction factors. Through ablation experiments comparing various weight settings, we determined that this ratio achieves a better balance between Area Under the Precision-Recall Curve (AUPRC) and Attack Success Rate (ASR), and thus is adopted as the default weighting scheme in FinSec’s Layer 4 calculations.

To systematically illustrate the overall structure of the FinSec model we developed, we provide an explanation as shown in Algorithm \ref{alg:finsec}. This process ensures that local risk estimates ($R_{\text{SAR}}$, $Q_t^{\text{def}}$, $\hat{y}$) are continuously calibrated and propagated within the system, thereby enabling precise identification of complex financial attacks within a unified risk space.

\begin{algorithm}[t]
\caption{FinSec: Multi-Layered Adversarial Risk Assessment}
\label{alg:finsec}
\SetAlgoLined
\DontPrintSemicolon

\KwIn{Context $C_t$, User $U_t$, History $H_t$, Thresholds $\tau_{\text{roll}}, \tau_{\text{block}}$.}
\KwOut{Final Risk $R_{\text{FIN}}(t)$, Decision $D_t$, Propagation $P_t$.}

\textbf{Phase 1: Pattern Detection (Layer 1)}\;
$M_k(X_{W_t}) \leftarrow \lambda_1 \text{Hit}_k + \lambda_2 \text{Sim}_k + \lambda_3 \text{Order}_k$\;
$R_{\text{SAR}}(t) \leftarrow \text{Aggregate}(\{M_k\}_k)$\;

\textbf{Phase 2: Deferred Risk Simulation (Layer 2)}\;
$Q_t^{\text{def}} \leftarrow f(R_{\text{SAR}}(t), U_t, H_t)$; \ $P_t \leftarrow 1 - e^{-\lambda Q_t^{\text{def}}}$\;
Initialize $R_{\text{ADV}}(t) \leftarrow Q_t^{\text{def}}$ and $\mathcal{S}_{adv} \leftarrow \emptyset$\;
\If{$Q_t^{\text{def}} \ge \tau_{\text{roll}}$}{
    $\mathcal{S}_{adv} \leftarrow \text{GenScenarios}(C_t \mid Q_t^{\text{def}})$\;
    $\text{DR}_{\text{roll}}(t) \leftarrow \frac{1}{N} \sum_{n=1}^{N} \max_k (\alpha' S_{t+k}^{(n)} + \gamma' \text{ADV}_{t+k}^{(n)})$\;
    $R_{\text{ADV}}(t) \leftarrow \max(Q_t^{\text{def}}, \text{DR}_{\text{roll}}(t))$\;
}

\textbf{Phase 3: Semantic Safety Analysis (Layer 3)}\;
$\hat{y} \leftarrow \text{LLM}_{\phi}(C_t, \mathcal{S}_{adv})$; \ $R_{\text{SEM}}(t) \leftarrow \text{ScoreMapping}(\hat{y})$\;

\textbf{Phase 4: Adaptive Fusion (Layer 4)}\;
\eIf{$\exists \text{Inj} \in \mathcal{S}_{adv} \lor \hat{y} = \text{Unsafe}$}{
    $\mathbf{w} \leftarrow [\alpha_{adv}, \beta_{adv}, \gamma_{adv}]^T$\;
}{
    $\mathbf{w} \leftarrow [\alpha_{base}, \beta_{base}, \gamma_{base}]^T$\;
}
$R_{\text{FIN}}(t) \leftarrow \alpha R_{\text{SAR}}(t) + \beta R_{\text{ADV}}(t) + \gamma R_{\text{SEM}}(t)$\;
$D_t \leftarrow \mathbb{I}(R_{\text{FIN}}(t) > \tau_{\text{block}})$\;

\KwRet $R_{\text{FIN}}(t), D_t, P_t, \mathcal{S}_{adv}$\;
\end{algorithm}

\section{Experiments and result}

\subsection{Original LLM selection}

\begin{figure*}[t]
    \centering
    \subfloat[Overall F1 Score]{
    \includegraphics[width=0.22\linewidth]{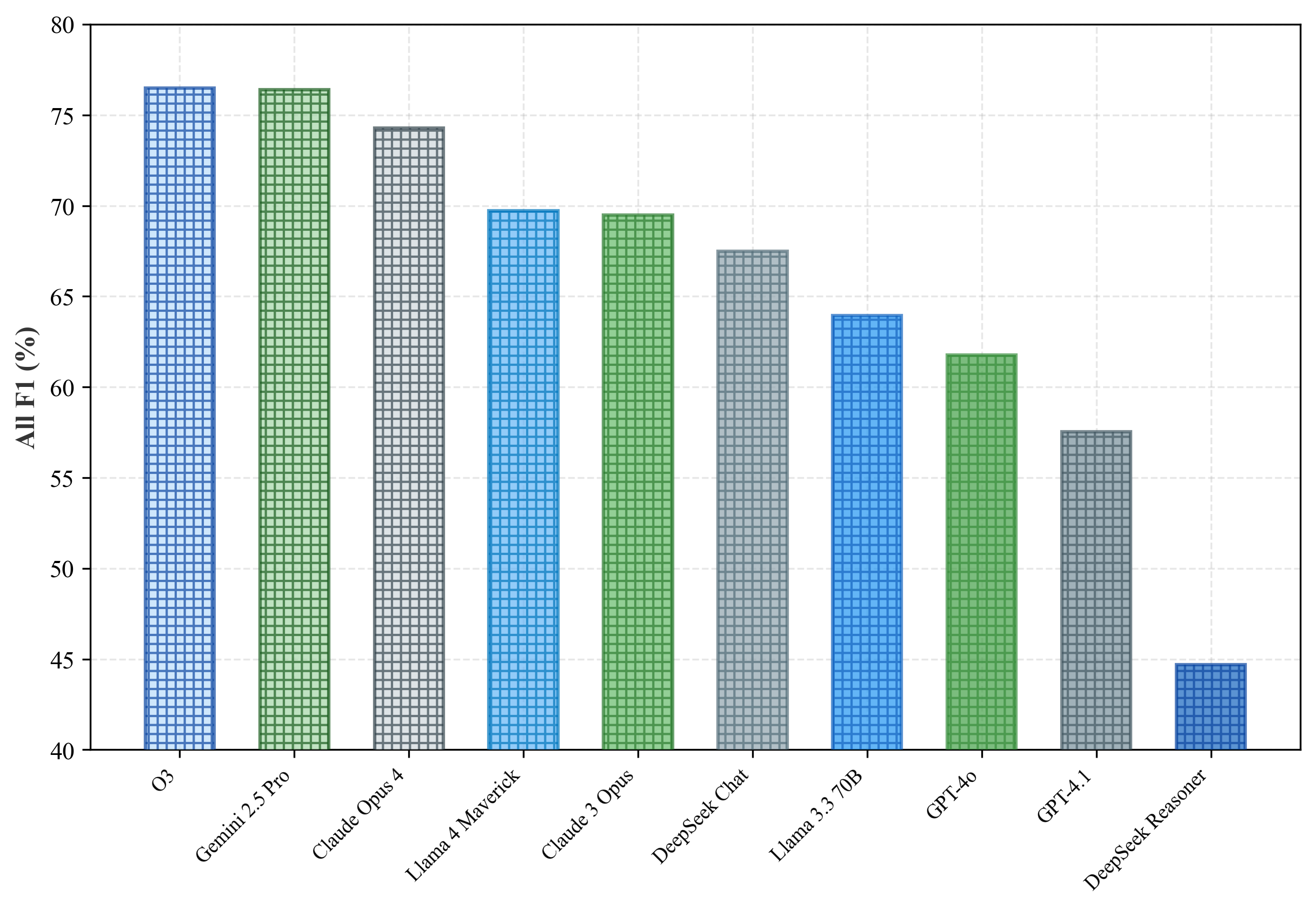}
        \label{fig:panel1}
    } \hfill
    \subfloat[Injection F1 Score]{
        \includegraphics[width=0.22\linewidth]{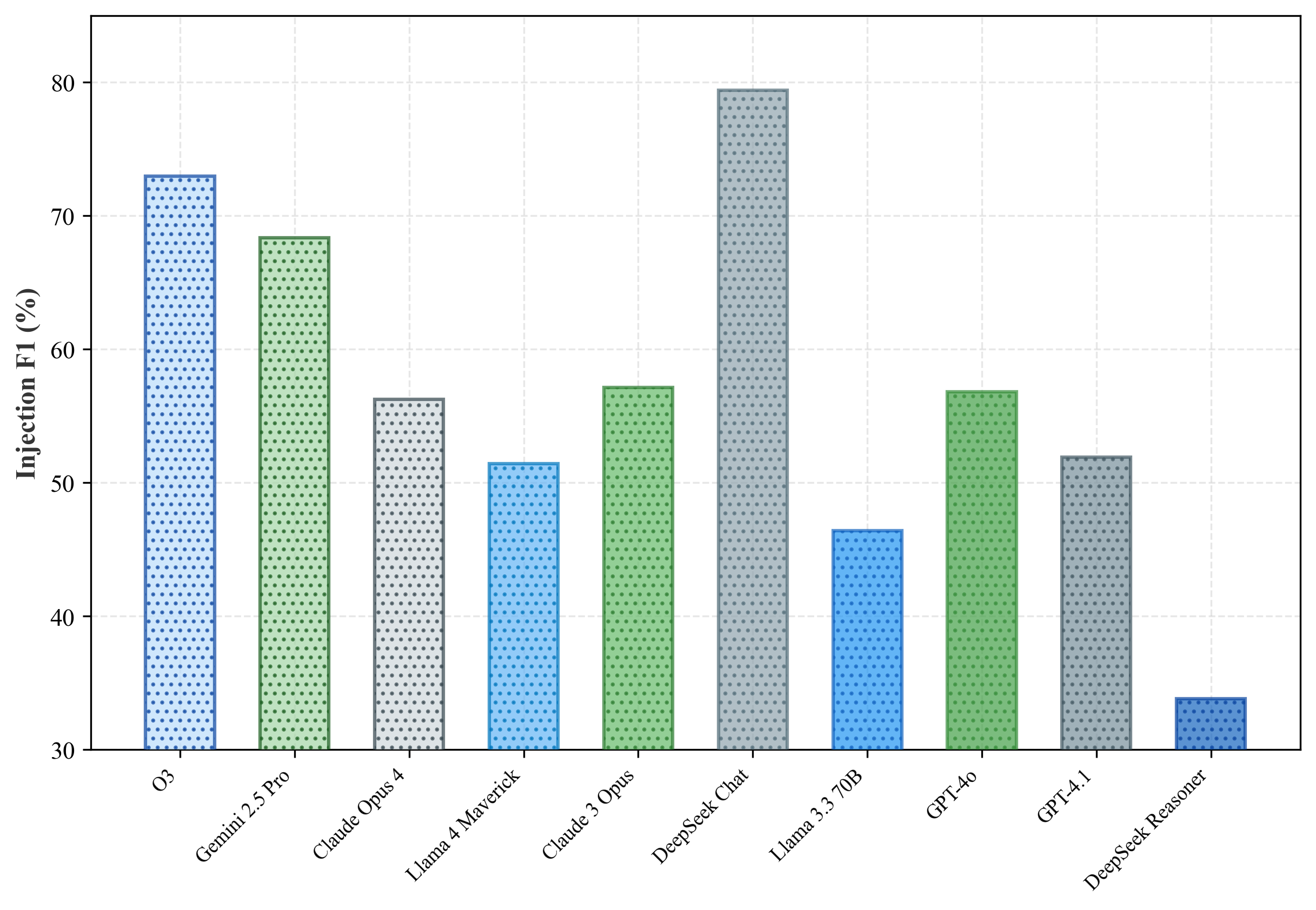}
        \label{fig:panel2}
    } \hfill
    \subfloat[Injection Recall]{
        \includegraphics[width=0.22\linewidth]{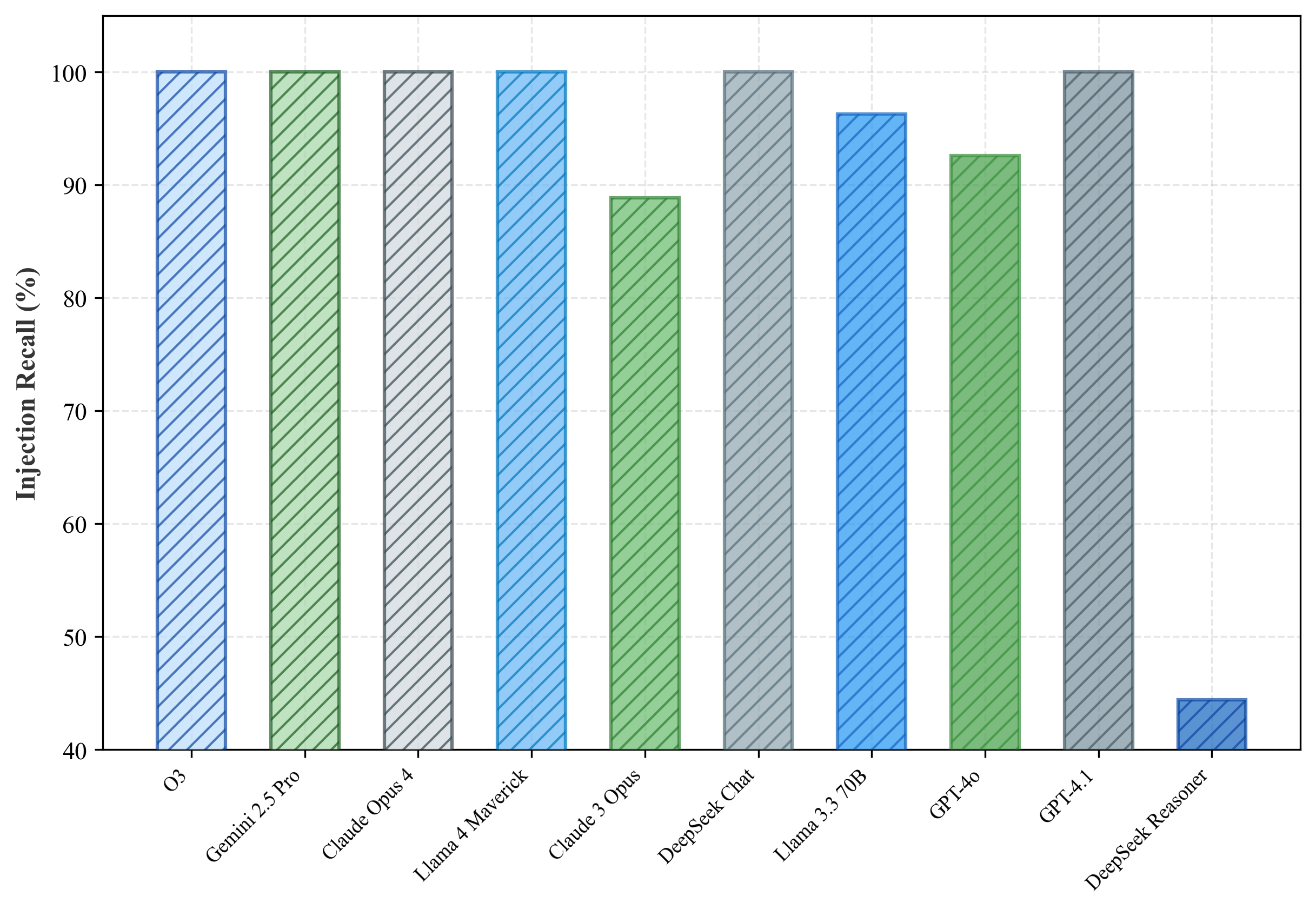}
        \label{fig:panel3}
    } \hfill
    \subfloat[Injection Specificity]{
        \includegraphics[width=0.22\linewidth]{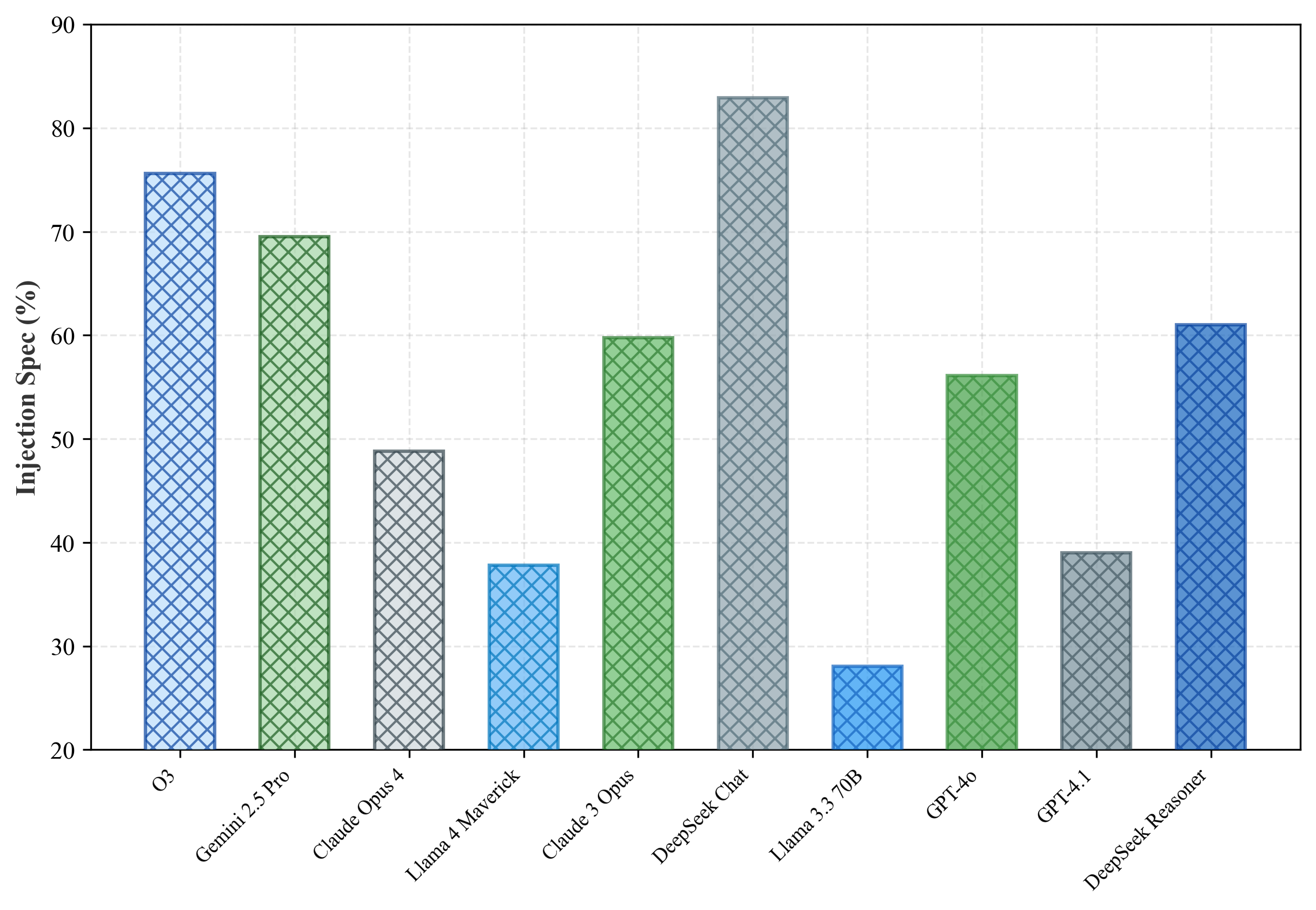}
        \label{fig:panel4}
    }
    \\ 
    \vspace{0.5em} 
    
    \subfloat[Unintended F1]{
        \includegraphics[width=0.22\linewidth]{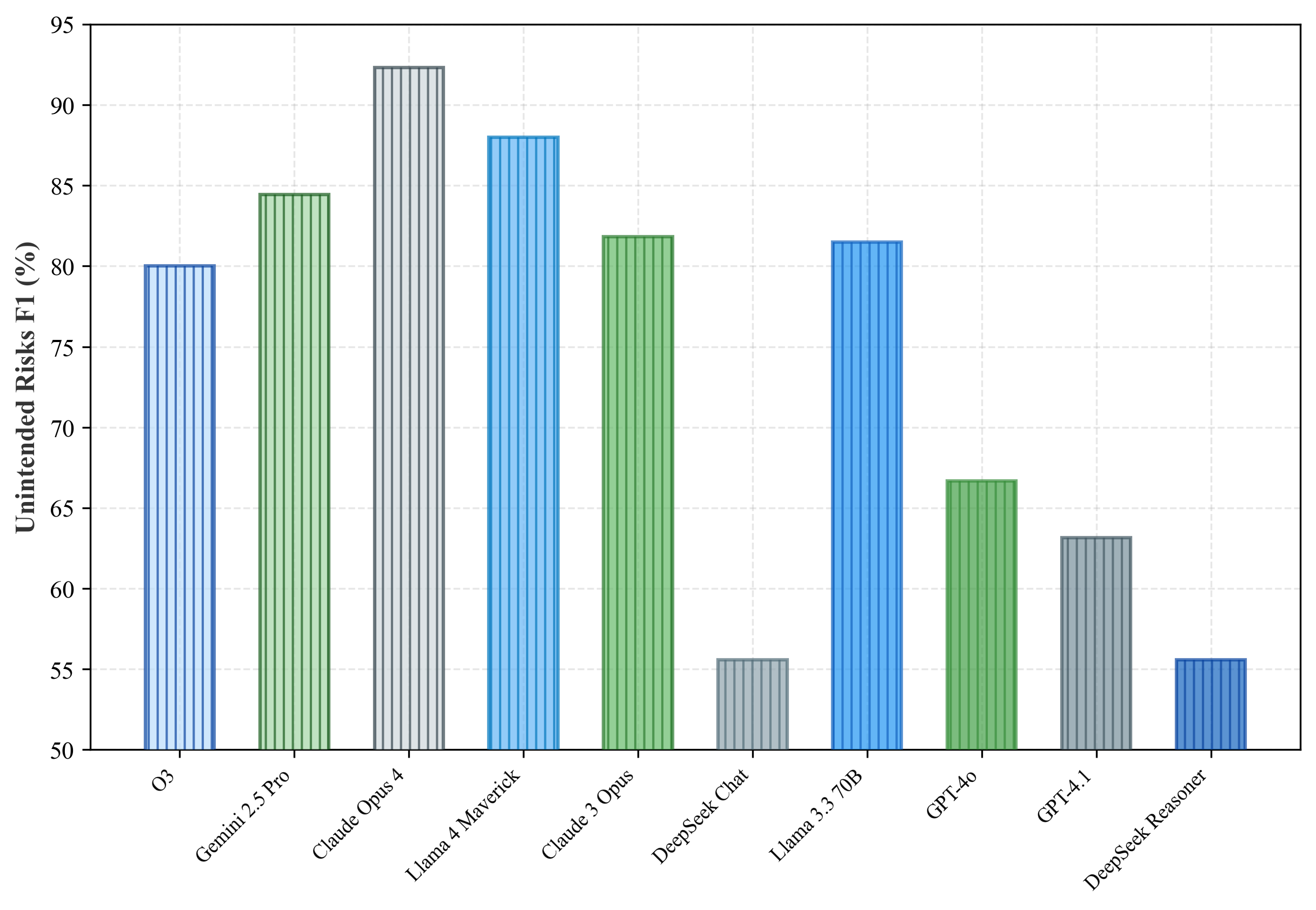}
        \label{fig:panel5}
    } \hfill
    \subfloat[Unintended Recall]{
        \includegraphics[width=0.22\linewidth]{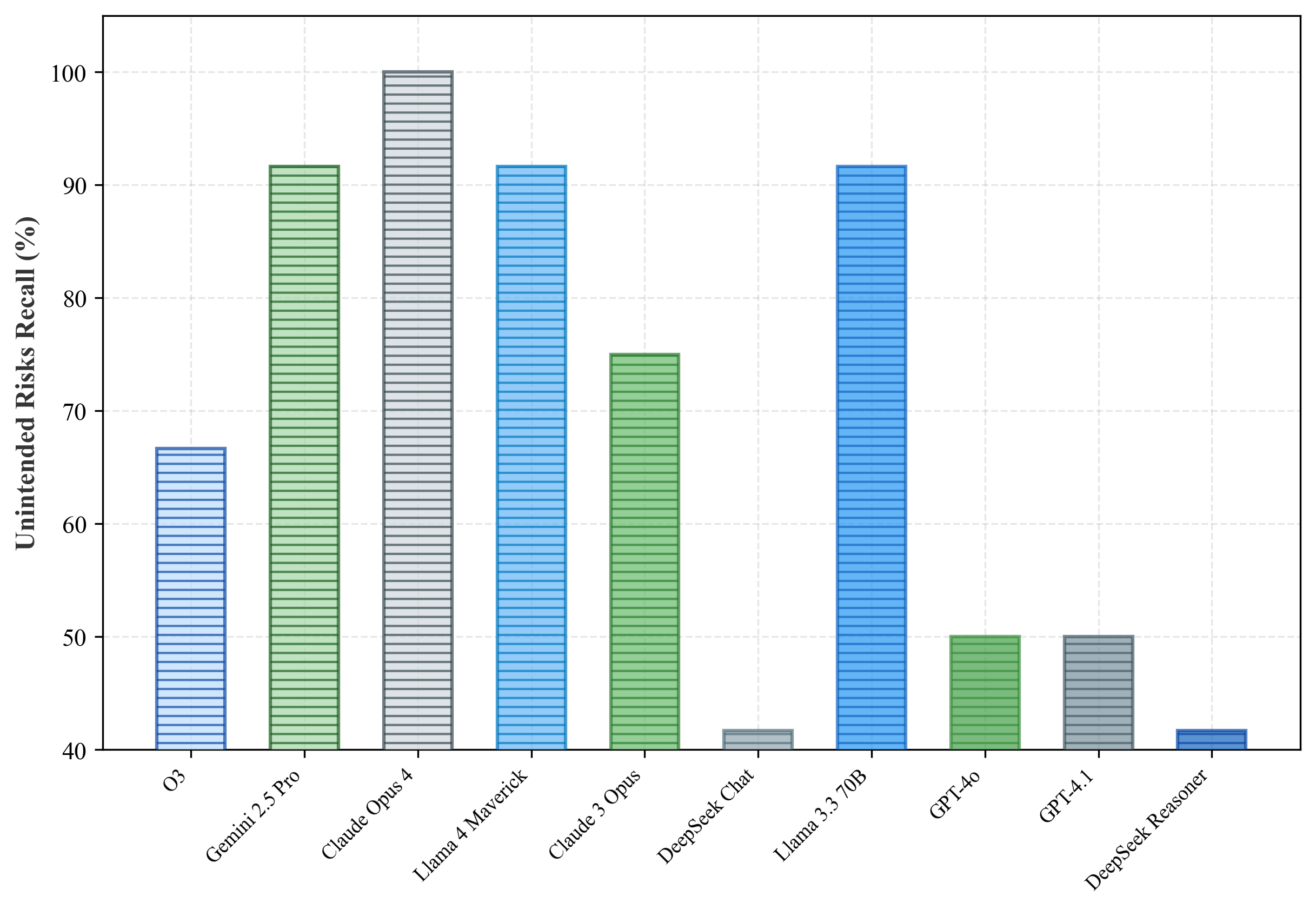}
        \label{fig:panel6}
    } \hfill
    \subfloat[Unintended Specificity]{
        \includegraphics[width=0.22\linewidth]{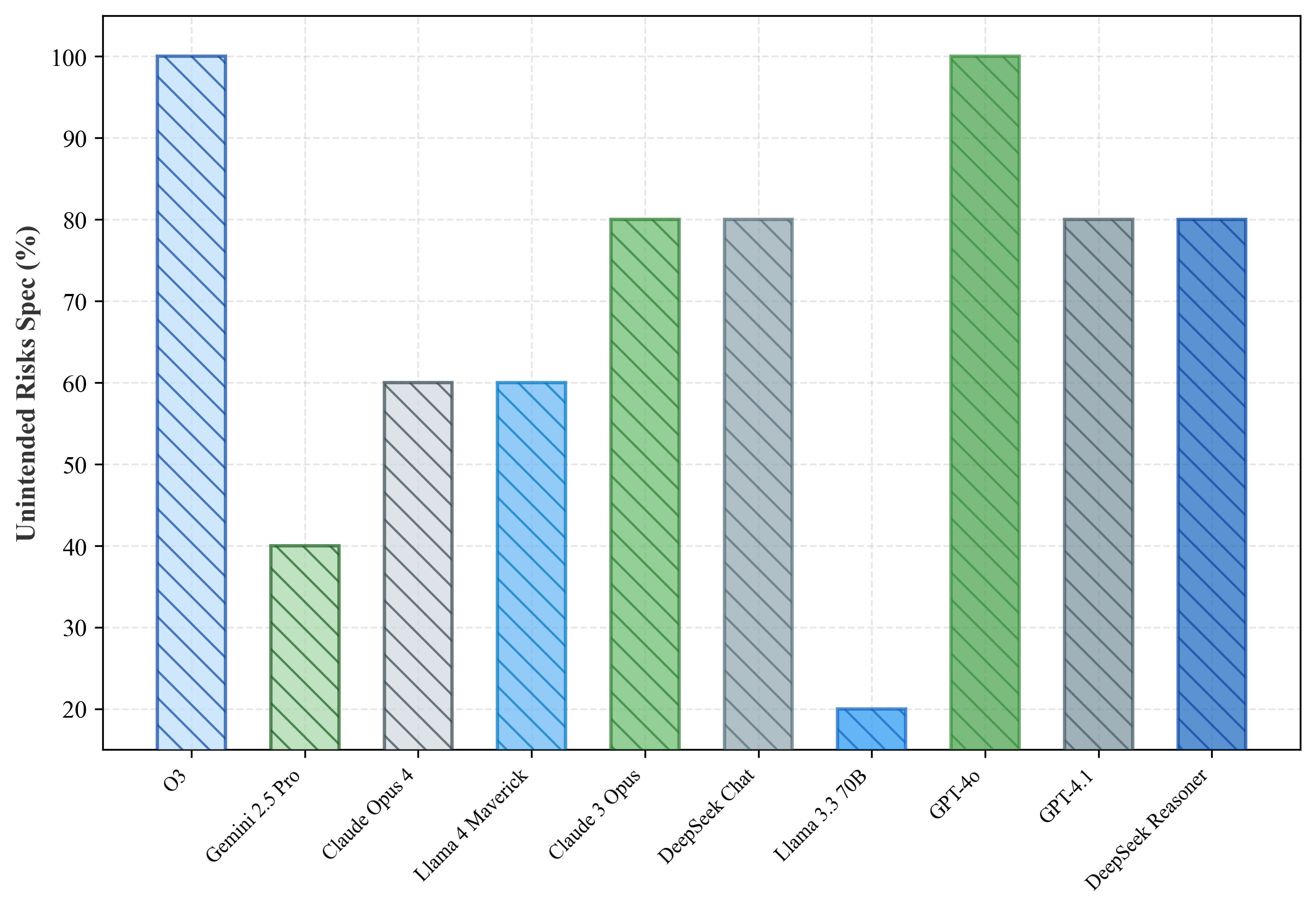}
        \label{fig:panel7}
    } \hfill
    \subfloat[Modified Sharpe Ratio]{
        \includegraphics[width=0.22\linewidth]{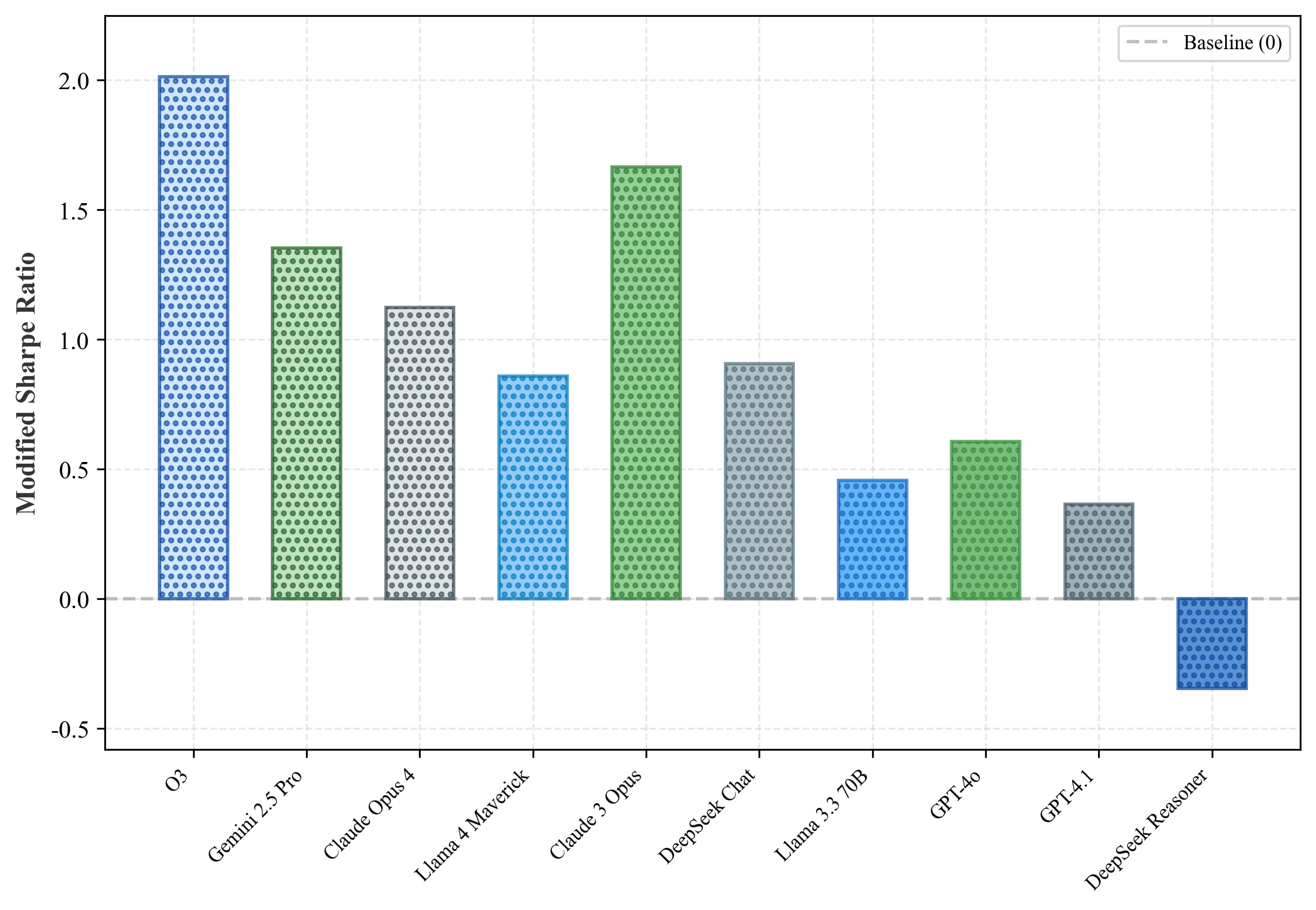}
        \label{fig:panel8}
    }
    
    \caption{Performance Comparison of 10 LLM Models on Financial Security Risk Assessment}
    \label{fig:eight_panels}
\end{figure*}

To comprehensively evaluate the benchmark performance of current mainstream large models in the field of financial security agent dialogues, we selected 10 advanced models and conducted extensive comparative experiments to assess their capability for financial agent security detection. We illustrate the specific comparative results across eight dimensions in Fig.~\ref{fig:eight_panels}. Our evaluation metrics include F1 score, recall, specificity, and a comprehensive modified sharpe ratio. We design a Sharpe-inspired performance index to measure risk-adjusted stability across evaluation metrics.

As shown in Fig.~\ref{fig:panel1}, we observe that O3 emerges as the leading model with an overall F1 score of 76.48\%, marginally outperforming Gemini 2.5 Pro (76.40\%). However, a granular examination highlights distinct performance profiles across categories. Regarding injection defense in Fig.~\ref{fig:panel2}, we find that DeepSeek Chat demonstrates superior efficacy (F1: 79.41\%), whereas other models struggle to balance recall and specificity (Fig.~\ref{fig:panel3} and Fig.~\ref{fig:panel4}). 
Regarding unintended risks, Claude Opus 4 adopted a highly conservative strategy, achieving perfect sensitivity (100.00\% recall) but significantly compromising specificity (48.78\%). Conversely, GPT-O3 maintained a more robust equilibrium between the two metrics (specificity: 75.61\%). These findings suggest that current SOTA models possess high baseline security. 
Finally, Fig.~\ref{fig:panel8} presents the modified sharpe ratio, where we evaluate the stability of model performance relative to risk, providing a robust metric for model reliability in volatile financial environments.
However, calibrating the trade-off between over-refusal and missed detections remains a critical optimization bottleneck, particularly in financial contexts.

\subsection{FinO3 }
We first optimized the prompts of large language models to meet the specific requirements of financial security detection, aiming to improve their specialization and accuracy in the domain of financial agent dialogue risk assessment. Subsequently, we integrated adversarial structures into the model prompts, enabling the models to more effectively identify potential attack behaviors and evasive language. This enhances their overall ability to detect security risks within financial agent dialogues. The experimental results for the FinO3 model, both before and after the inclusion of adversarial structures (FinO3adv), are presented in Table~\ref{tab:o3_fino3_comparison}.


\begin{table*}[!t]
\centering
\caption{Comparison of R-judge(O3) and FinO3 Variants on Financial Security Risk Assessment }
\renewcommand{\arraystretch}{1.15}
\setlength{\tabcolsep}{6pt}
\begin{tabular}{l|c|ccc|ccc}
\hline
\multicolumn{1}{c|}{\textbf{Models}} & \multicolumn{1}{c|}{\emph{All}} &
\multicolumn{3}{c|}{\emph{Injection Attacks}} &
\multicolumn{3}{c}{\emph{Unintended Risks}} \\
\cline{2-2}\cline{3-5}\cline{6-8}
& \textbf{F1 (\%)} &
\textbf{F1 (\%)} & \textbf{Recall (\%)} & \textbf{Spec (\%)} &
\textbf{F1 (\%)} & \textbf{Recall (\%)} & \textbf{Spec (\%)} \\
\hline
R-Judge(O3)                & 76.49 & 80.00 & 66.67  & \textbf{100.00} & 72.97 & 100.00 & 75.61 \\ \cline{1-8}
FinO3 (Zero-shot) & 82.05 & 86.96 & 83.33  & 80.00          & 77.14 & 100.00 & 80.49 \\ \cline{1-8}
FinO3 (Few-shot)  & 83.27 & 73.68 & 63.64  & 66.67          & \textbf{92.86} & \textbf{100.00} & \textbf{95.06} \\ \cline{1-8}
FinO3adv & 82.85 & 83.87 & 100.00  & 87.65         & 81.82 & 81.82 & 33.33 \\ \cline{1-8}

\hline
\end{tabular}
\label{tab:o3_fino3_comparison}
\end{table*}

In the case of FinO3adv, although adversarial verification structures were incorporated into the prompt, the overall model performance did not improve significantly, and its ability to identify unintended attacks even declined. This indicates that simply introducing complex adversarial detection mechanisms into the prompt can create tension between defensive strength and the original task objectives, resulting in model confusion when faced with covert attacks and, consequently, reduced detection accuracy.
Specifically, integrating adversarial logic at the prompt level works well when the detection architecture has only a single layer. However, as the detection model becomes more complex, the experimental results indicate that this method is much less effective.

This is primarily because embedding adversarial structures within the prompt can lead to excessive defensiveness in the model. In particular, since the task requires detecting attacks while also performing classification, the model faces a trade-off between these objectives. This tension ultimately results in suboptimal performance. Such conflicts are especially pronounced in financial dialogue scenarios, as the context typically involves sensitive terms such as amounts, accounts, and permissions. If the model’s defensiveness is set too high, ordinary transaction requests may be misclassified as attacks.

Therefore, we observed that the delayed manifestation of adversarial semantics is quite apparent in financial dialogues. The prompt layer, lacking temporal sensitivity, can only constrain the semantics of the current turn, rendering it ineffective in modeling the evolving risks of dialogue over time. Consequently, when further developing FinSec, we adjusted the placement of the adversarial model, relocating it from the third layer. The third layer now serves independently as a semantic security discrimination layer.

\subsection{FinSec}
\subsubsection{Layer 1 and layer 2}

In layer 1, following the guidelines of AML and SAR, we introduce six primary detection patterns: structuring, gradual extraction, rapid escalation, circular flow, timing anomaly and social engineering. Among the broad spectrum of financial activities, each suspicious pattern is assigned a different weight depending on the type of financial operation. To align with the overall direction of market regulation, we assign weights ti each pattern.

In the delay risk simulation module of layer 2, we integrate the calculation of abnormal risk scores from multiple risk factors according to the characteristics of the financial market. These factors include user history, transaction patterns, amount anomalies, frequency anomalies, and timing anomalies. For the current dataset, frequency anomaly demonstrates the most significant discrimination ability. Therefore, we evaluate the weight assignment based on a multidimensional assessment framework across four dimensions: discrimination, balance, robustness, and scalability.

\begin{figure}[t]
    \centering
    \includegraphics[width=\columnwidth]{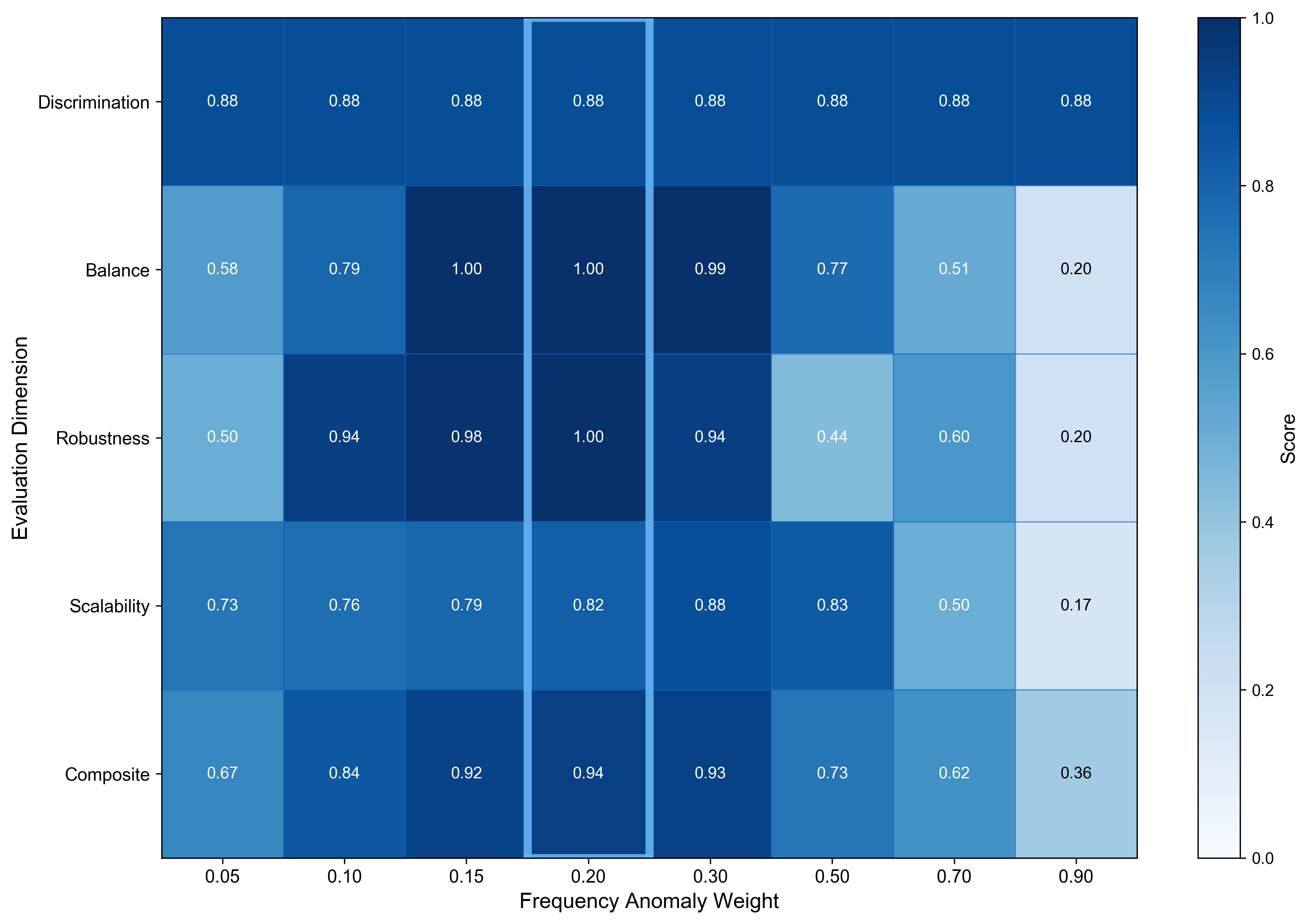}
    \caption{Global view of performance stability via score heatmap}
    \label{fig:L21}
\end{figure}

\begin{figure}[t]
    \centering
    \includegraphics[width=\columnwidth]{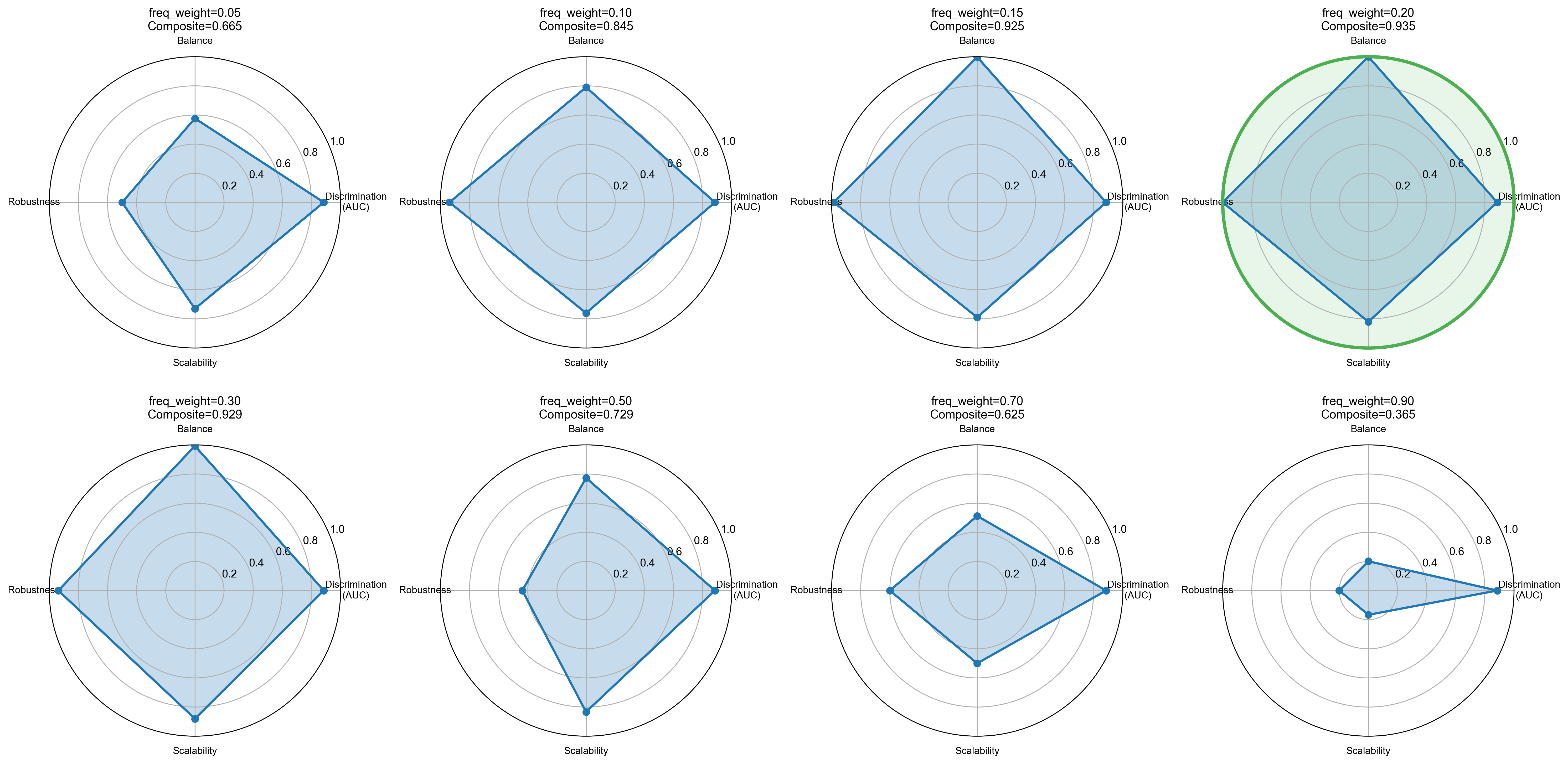}
    \caption{Trade-off analysis and capability balance under different weights}
    \label{fig:L22}
\end{figure}

\begin{figure}[t]
    \centering
    \subfloat[Impact of weight variation on individual performance metrics\label{fig:L231}]{%
        \includegraphics[width=0.45\linewidth]{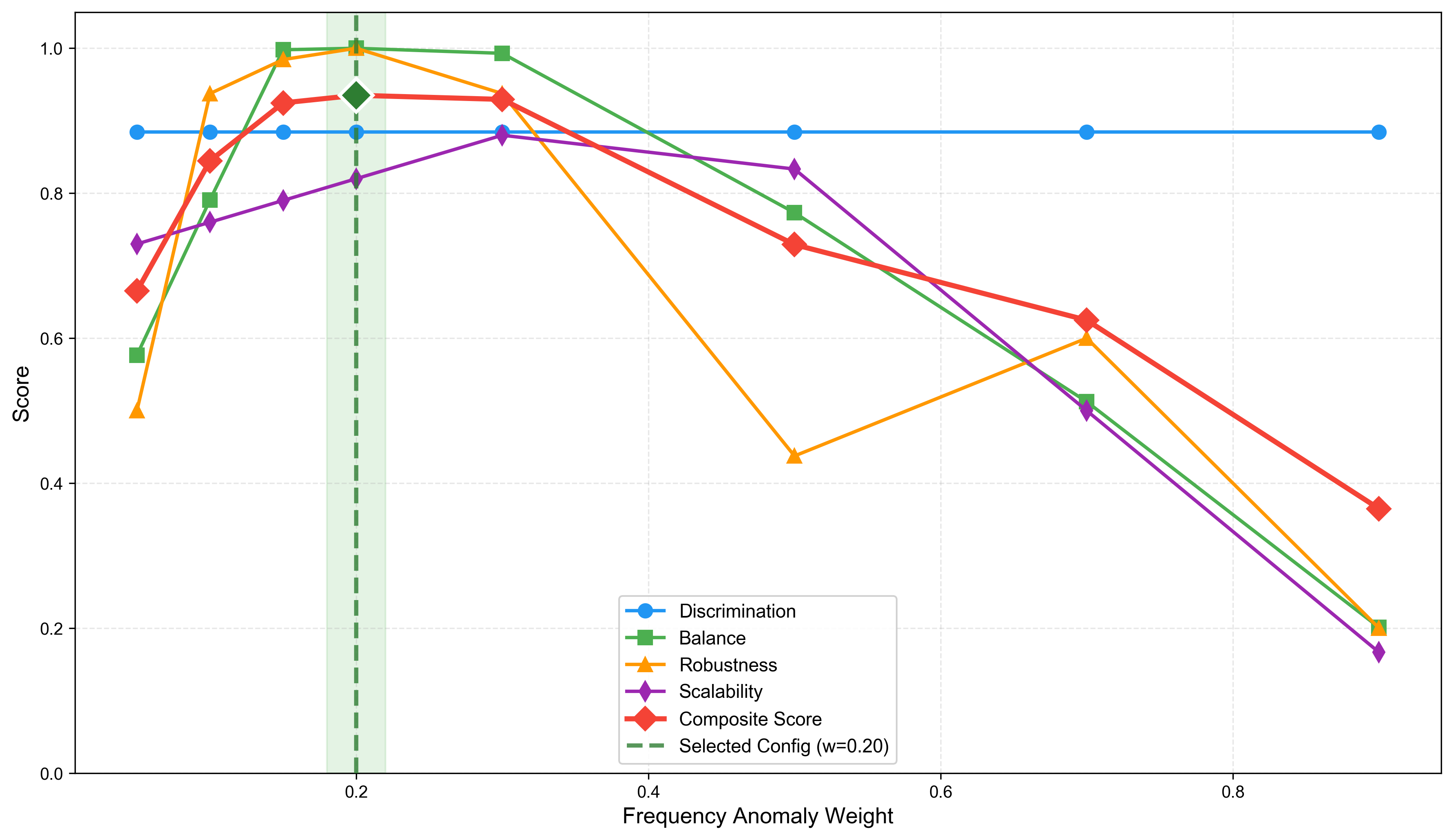}
    }
    \hfill
    \subfloat[Final selection based on aggregated composite scores
    \label{fig:L232}]{%
        \includegraphics[width=0.45\linewidth]{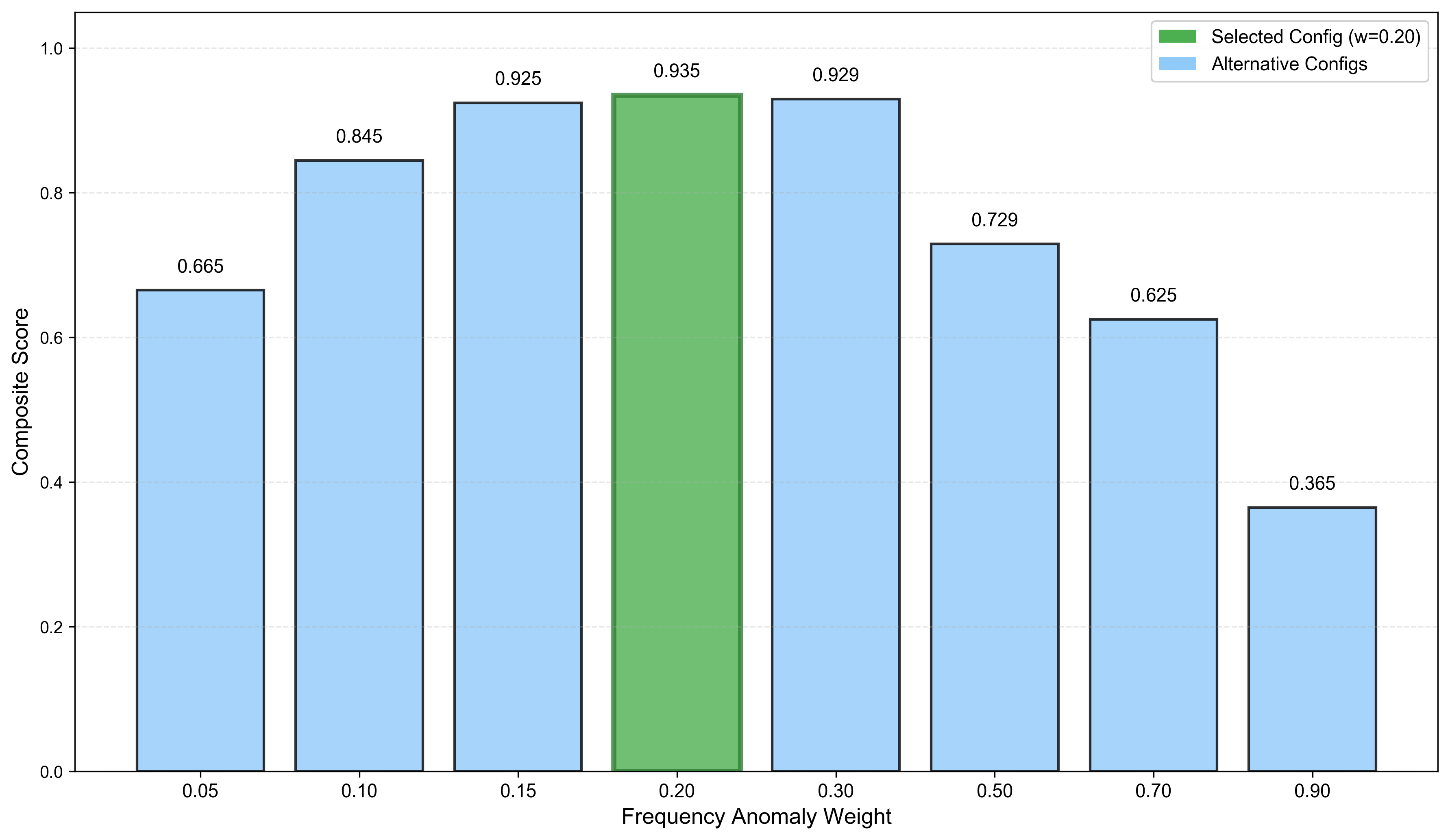}
    }
    \caption{Performance Sensitivity and Optimal Weight Determination for Frequency Anomaly Detection}
    \label{fig:L23}
\end{figure}

As shown in Fig.~\ref{fig:L21}  displays the scores of each configuration across Discrimination, Balance, Robustness, and Scalability. The configuration with weight \( w = 0.2 \) performs well in all four dimensions with no obvious weaknesses. Fig.~\ref{fig:L22}  shows the score distribution of different weight configurations for each dimension, helping to identify the optimal interval. At w=0.20, both Balance and Robustness achieve full marks, and the Composite score reaches its peak. Fig.~\ref{fig:L231}  illustrates the curves of the four-dimensional scores and the composite score as the weight changes, allowing identification of the optimal weight range. The composite score maintains a high level in the range of \( w = 0.15-0.3 \), peaking at \( w = 0.2 \). Fig.~\ref{fig:L232}  is a bar chart comparing the composite scores of different weight configurations, clearly showing that the configuration with \( w = 0.2 \) is optimal. Based on this analysis, we determine the weight configuration for frequency anomalies. In this way, current performance and future adaptability are balanced, allowing sufficient weight for effective factors when processing data with new risk variables (such as user histories) in the future. At the same time, the system’s robustness against measurement errors is also improved.

\subsubsection{Overall Model Results Validation }

To evaluate the effectiveness of FinSec in identifying anomalous and risky dialogues within financial agent conversations, we introduce the Area Under the Precision–Recall Curve (AUPRC) and the Attack Success Rate (ASR) as metrics\cite{5.1}\cite{5.2}. The formulas are presented in Eq.\eqref{eq:auprc} and Eq.\eqref{eq:asr} as follows:
\begin{equation}
\mathrm{AUPRC}_{\text{FinSec}}
= 
\sum_{q \in \{\mathrm{inj}, \mathrm{unint}\}}
w_q 
\int_0^1
\left(
\sum_{\ell=1}^{3}
\omega_\ell \, P_{q, L_\ell}(R)
\right)
\, dR
\label{eq:auprc}
\end{equation}

Standard AUPRC models lack the hierarchical granularity required to effectively evaluate FinSec. To address this limitation, we extended the traditional metric into a Structured AUPRC. This framework operates across three levels, calculating performance independently for two distinct attack vectors. Within each level, risk contributions are weighted according to FinSec’s hierarchical architecture. Specifically, $q$ denotes the attack type (categorized as injection or unintended), and $P_{q, L_{\ell}}(R)$ represents the precision at layer $\ell$ for attack type $q$ at a given recall threshold $R$.

\begin{equation}
\mathrm{ASR}_{\text{FinSec}}
=
1 -
\sum_{q \in \{\mathrm{inj}, \mathrm{unint}\}}
w_q
\sum_{\ell=1}^{3}
\eta_{q,\ell}
\frac{\mathrm{TP}_{q, L_\ell}}{|\mathcal{A}_q|}
\label{eq:asr}
\end{equation}

Meanwhile, we established a hierarchical robust ASR for FinSec. Mirroring the logic described above, this metric incorporates weighting based on the model's architecture and risk profile. Specifically, after determining the count of successfully identified attacks at layer $\ell$ (denoted as $TP_{q,L_{\ell}}$), we perform a second-order aggregation across both attack types and hierarchical layers."
Ultimately, we obtain a comprehensive security score that accounts for detection performance, adversarial robustness, and output stability. The specific formulation is given in equation \eqref{eq:Rfinsec}.

\begin{equation}
\mathcal{R}_{\text{FinSec}}
=
\alpha \, \mathrm{AUPRC}_{\text{FinSec}}
+
\beta \left( 1 - \mathrm{ASR}_{\text{FinSec}} \right)
+
\gamma \, \mathrm{Stab}_{\text{FinSec}}
\label{eq:Rfinsec}
\end{equation}

    

\begin{table*}[t]
\centering
\caption{F1-Score Comparison of Financial Security Risk Assessment Models}
\label{tab:final_f1}
\renewcommand{\arraystretch}{1.15}
\setlength{\tabcolsep}{6pt}
\begin{tabular}{l|c|c|c}
\hline
\multicolumn{1}{c|}{\textbf{Models}} & \multicolumn{1}{c|}{\textbf{Utility F1 (\%)}} &
\multicolumn{1}{c|}{\textbf{Injection Attack F1 (\%)}} &
\multicolumn{1}{c}{\textbf{Unintended Attack F1 (\%)}}\\
\hline
O3 (R-Judge) & 76.48 & 72.97 & 80.00 \\ \cline{1-4}
FinO3 & 76.48 & 83.27 & 73.68 \\ \cline{1-4}
FinO3adv & 82.94 & 83.87 & 81.82 \\ \cline{1-4}
FinSec & \textbf{90.13} & \textbf{94.55} & \textbf{85.71} \\ \cline{1-4}
\hline
\end{tabular}
\end{table*}

Table\ref{tab:final_f1} compares teh effectiveness of R-Judge, FinO3, FinO3adv, and FinSec.
Specifically, regarding injection attack detection, compared to the baseline R-Judge (72.97\%), FinSec demonstrates significant improvement. This result indicates that FinSec is highly effective in handling proactive malicious command attacks.
The nest column illustrates the detection of unintended risk. Here, FinSec continues to deliver leading performance (85.71\%). Notably, although R-Judge achieves reasonable results (80.00\%), FinO3 exhibits a noticeable decline in performance (73.68\%). This suggests that incorporating adversarial detection methods into the prompt can lead to degraded performance under poisoning attacks, possibly due to excessively long prompt texts.
Finally we examine the overall F1 scores. FinSec is the only model to achieve a score  exceeding 90\%, maintaining high precision while ensuring a high recall rate. It achieves substantial improvements in both detection ability and precision, demonstrating stable performance in adversarial scenarios, and providing robust and balanced defense against both types of financial dialogue risks.

\begin{figure}[!h] 
    \centering
    \subfloat[Defense Rate with AUPRC \label{fig:asr1}]{%
        \includegraphics[width=0.49\linewidth]{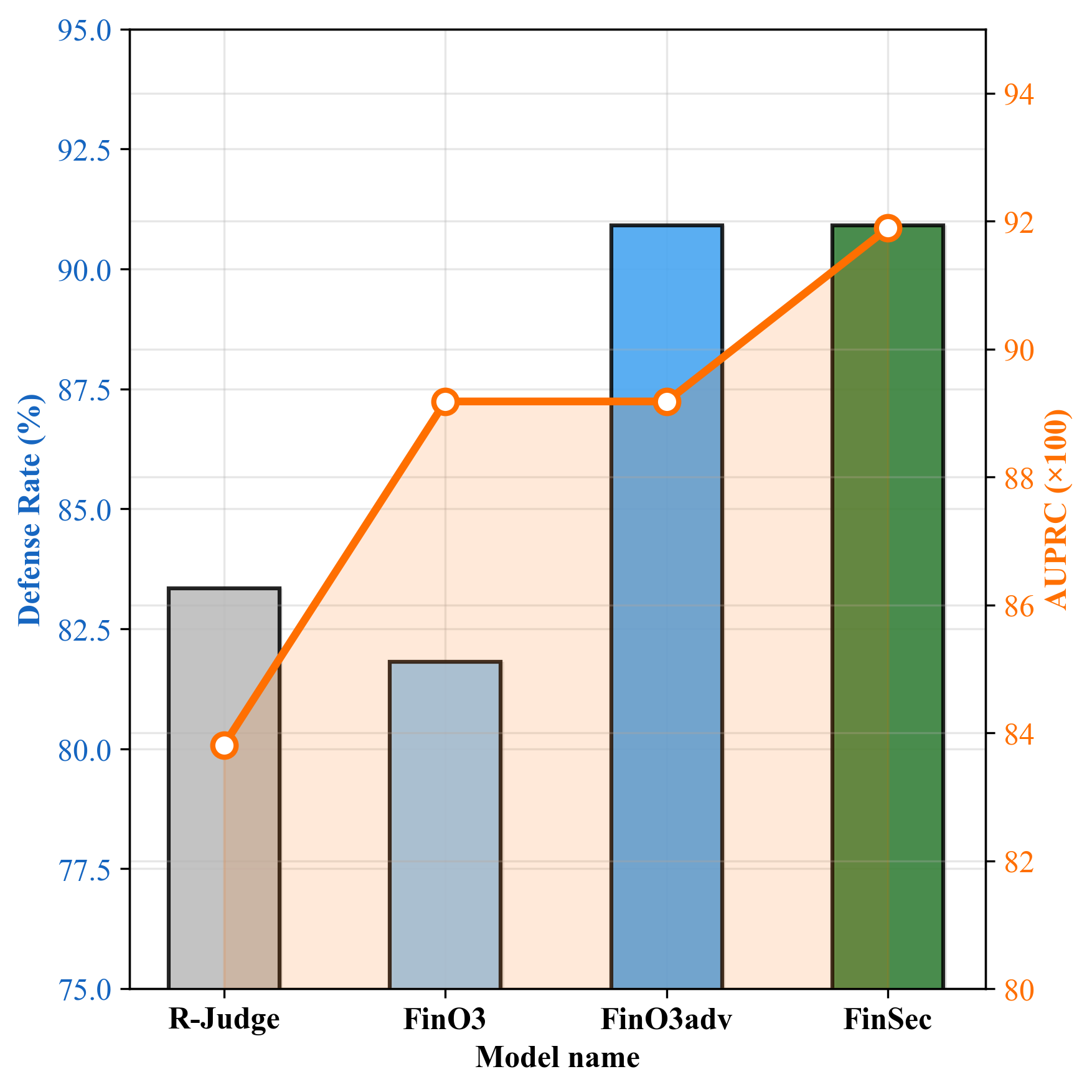}}
    \hfill 
    \subfloat[Comprehensive Score and ASR \label{fig:asr2}]{%
        \includegraphics[width=0.49\linewidth]{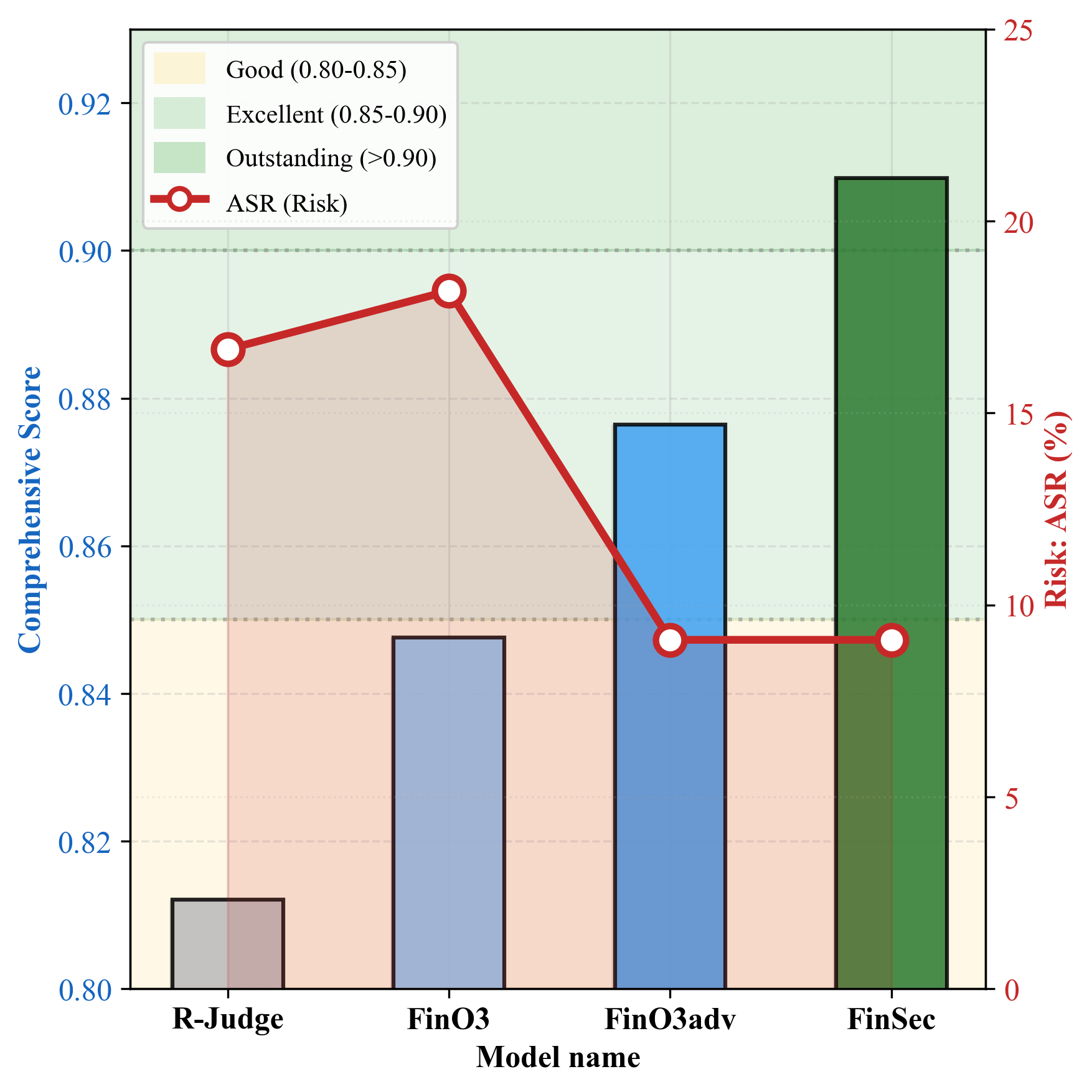}}
    
    \caption{Performance Evaluation: Trade-off Analysis and Final Verdict:
     (a) Robustness Analysis: Defense Rate and AUPRC; 
     (b) Final Verdict: Comprehensive Score versus Risk (ASR)}
    \label{fig:asr} 
\end{figure}

Figure~\ref{fig:asr} shows the analyzes of model performance from a multi-dimensional trade-off perspective. First, we evaluate the robustness of each model in Fig.~\ref{fig:asr}(a), specifically regarding both defense capabilities and precision. Here, the bar chart displays the defense rate, while the red line corresponds to the AUPRC. Through these results, we confirm the effectiveness of our proposed defense mechanisms, which maintain strong performance even on imbalanced datasets.

Next, we present the final verdict in Fig.~\ref{fig:asr}(b), contrasting the comprehensive score against the potential risk (ASR). We observe that FinSec achieves a comprehensive weighted score of 0.9098, significantly outperforming the comparative models. Notably, compared with the baseline R-Judge model, we achieve a 12\% overall performance improvement while maintaining the lowest attack success rate (indicated by the red line). This demonstrates that FinSec ensures superior utility without compromising the security of agent dialogues.

\begin{figure}[t]
    \centering
    \includegraphics[width=\columnwidth]{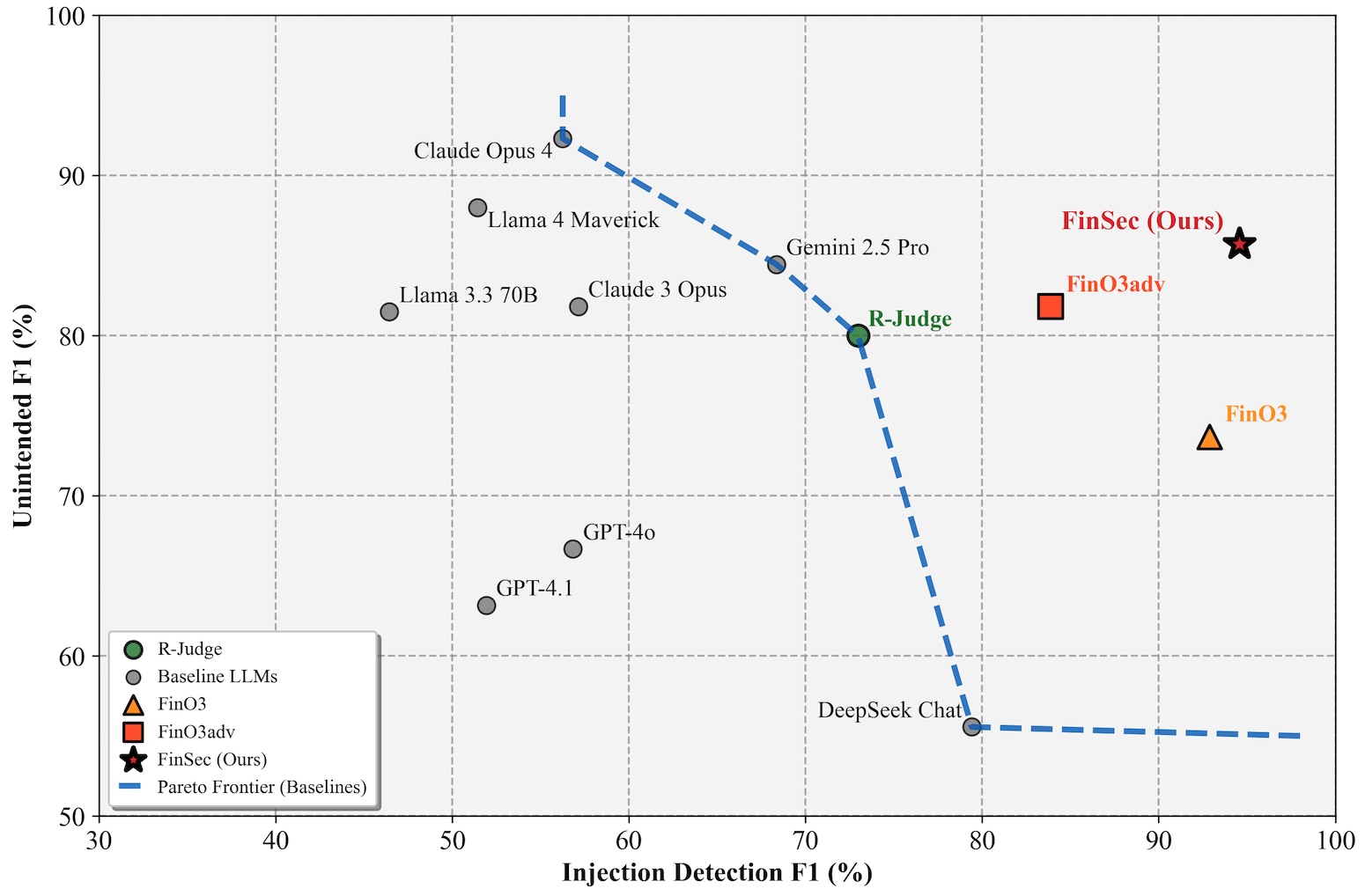}
    \caption{Performance Trade-off: Injection and Unintended Risks}
    \label{fig:finall}
\end{figure}

Figure ~\ref{fig:finall} presents the results of a Pareto frontier analysis for various models on injection risk and unintended risk detection tasks in financial agent dialogues. As illustrated, the blue dashed line delineates the performance boundaries of leading LLMs such as Claude Opus4 and Gemini2.5 Pro. A clear trade-off phenomenon is observed: most models excel in only one dimension. However, our proposed model, FinSec, breaks through this research bottleneck. Its performance significantly surpasses the Pareto frontier formed by baseline models. This demonstrates that FinSec’s unique hierarchical defense architecture avoids the typical zero-sum trade-off, successfully combining strong robustness against malicious instructions with high sensitivity to semantic risks, thereby achieving dual optimality in financial agent dialogue scenarios.

\section{Conclusions}
This paper introduces a hierarchical detection framework for financial agent dialogues security, named FinSec. FinSec employs a three-layer architecture to evaluate the safety of financial agent dialogues from multiple perspectives, producing an integrated security assessment as the final output. FinSec addresses critical limitations of existing LLMs, which struggle to balance high sensitivity and operational practicality in financial environments. Operational practicality refers to the system’s ability to efficiently accomplish legitimate financial security assessment tasks while maintaining compliance and safety. FinSec enhances the precision of intent recognition and, importantly, avoids misclassifying harmless instructions as attacks.

Methodologically, Layer 1 of FinSec constructs a behavioral pattern library based on international AML/SAR standards and employs a “triple matching” mechanism for high-recall preliminary screening of explicit risks. Building upon this, Layer 2 quantifies delayed risks and utilizes adversarial detection to quantitatively evaluate the effectiveness of defense mechanisms. Layer 3 incorporates a semantic discriminator for security assessment, and finally, Layer 4 performs a comprehensive scoring and makes the final decision.
The results demonstrate that FinSec achieves state-of-the-art performance, significantly outperforming baseline models. Notably, in Pareto frontier analyses, FinSec effectively overcomes the specificity–recall bottleneck, mitigating the trade-off between excessive rejection and missed detections, thereby greatly enhancing the reliability of financial agent security detection.

Regarding potential improvements, we first observed that when prompts become excessively long or complex—i.e., under long-sequence input conditions—the outputs of large language models exhibit deviations, resulting in decreased accuracy. This effect is more pronounced compared to scenarios involving concise prompts. Additionally, as the overall model architecture becomes more complex, the inability of LLMs to effectively process large volumes of input at once further undermines accuracy. Therefore, it is necessary to optimize financial risk detection capabilities specifically for long-sequence prompts.
Secondly, in practical scenarios, major financial losses are often caused by intentional injection attacks. As a result, it is essential to refine the categorization of injection attacks and develop more targeted identification and defense mechanisms. In future work, we will continue to address these two outstanding issues through dedicated research.

\section{Acknowledgement}
This work was supported in part by the JSPS KAKENHI under Grants
23K11072, in part by the National Natural Science Foundation of China under Grants U21B2019 and 61972255.


%





\ifCLASSOPTIONcaptionsoff
  \newpage
\fi

{\footnotesize
\bibliographystyle{IEEEtran}
\bibliography{ref}
}

%




\end{document}